\documentclass{aastex63}
\usepackage[T1]{fontenc}
\def\simgr{\,\hbox{\hbox{$ > $}\kern -0.8em \lower 1.0ex\hbox{$\sim$}}\,}
\def\simle{\,\hbox{\hbox{$ < $}\kern -0.8em \lower 1.0ex\hbox{$\sim$}}\,}

\shortauthors{THORSTENSEN ET AL.}
\newcommand{\numberOfTargets}{8\ }
\shorttitle{\numberOfTargets Magnetic CVs}

\begin{document}
\title{Optical Studies of \numberOfTargets AM Herculis-Type 
Cataclysmic Variable Stars 
}

\author[0000-0002-4964-4144]{John R. Thorstensen
}

\affil{Department of Physics and Astronomy,
6127 Wilder Laboratory, Dartmouth College,
Hanover, NH 03755-3528}

\author{Mokhine Motsoaledi}
\affil{Department of Astronomy, University of Cape Town, Private Bag X3, Rondebosch 7701, South Africa}

\affil{South African Astronomical Observatory, PO Box 9, Observatory 7935, Cape Town, South Africa}
\author[0000-0002-6896-1655]{Patrick A. Woudt}
\affil{Department of Astronomy, University of Cape Town, Private Bag X3, Rondebosch 7701, South Africa}

\author{David A. H. Buckley}

\affil{South African Astronomical Observatory, PO Box 9, Observatory 7935, Cape Town, South Africa}

\author{Brian Warner}
\affil{Department of Astronomy, University of Cape Town, Private Bag X3, Rondebosch 7701, South Africa}

\begin{abstract}
We report detailed follow-up observations of \numberOfTargets cataclysmic
variable stars (CVs) that are apparently AM Her stars, also called
polars.  For all, we either determine orbital periods
for the first time, or improve on existing determinations.
The seven for which we have spectra show the high-amplitude radial 
velocity curves and prominent 
\ion{He}{2} $\lambda$4686 emission lines characteristic of 
strongly magnetic CVs, and their periods, which range from
81 to 219 minutes, are also typical for AM Her stars. 
Two objects from the Gaia-alerts
index, Gaia18aot and Gaia18aya, are newly identified
as CVs. Another, RX J0636.3+6554, eclipses deeply, while 
CSS080228:081210+040352 shows a sharp dip that is apparently a partial
eclipse. The spectrum of Gaia18aya has a cyclotron 
harmonic near $\lambda 5500$ \AA\ that constrains
the surface field to $\sim 49$ MG or greater.

\end{abstract}


\section{Introduction}

Cataclysmic variable stars \citep{warner95} are close binary 
systems in which a 
white dwarf accretes material from a more extended companion,
usually resembling a main-sequence star, which overflows its 
Roche lobe (critical equipotential surface). 
The name arose because the first known
examples underwent outbursts -- classical nova explosions
occur when nuclear fuel accumulated on the white dwarf's surface 
explodes, and the more common dwarf novae undergo outbursts when gas 
accumulated in an accretion disk becomes unstable, 
and rapidly accretes onto the white dwarf. 

Magnetic CVs -- in which the white dwarf is strongly magnetized 
-- can behave quite differently. They are usually much stronger X-ray emitters than non-magnetic CVs.
If the magnetic field is not especially strong, an accretion disk 
can form far from the white dwarf, but the inner disk is 
disrupted and the field forces material to fall onto the poles 
of the white dwarf.  Systems of this kind are called  
DQ Herculis stars \citep{patterson94}, or intermediate polars, and they
show pulsations in the X-ray and optical bands
at the rotation period of the white dwarf and/or the orbital sidebands (e.g.~the orbital-spin beat).  A still
stronger magnetic field can disrupt the formation of an accretion 
disk entirely; often, magnetic torques
force the white dwarf to co-rotate with the orbit, though
the coupling is weak enough that the 
white dwarf can sometimes be temporarily knocked out of co-rotation.
In these systems, at least some of the
matter lost from the companion threads onto the
magnetic field and falls directly onto the white dwarf's
magnetic poles via magnetically confined accretion columns.  CVs of this kind are classified 
as {\it AM Herculis stars}, after their prototype, and
are also called {\it polars} \citep{cropper90}, because they often
show strong circular polarization modulated at the 
the orbital (= rotational) frequency.


It is often easy to recognize an AM Her star 
even without polarization measurements.
When they are accreting actively, their spectra
show strong emission lines, with high excitation; 
the \ion{He}{2} $\lambda$4686 emission is usually comparable in
strength to H$\beta$.  Because the emission lines arise largely
in the accretion column, their radial velocities are
often dominated by infall, which can reach velocities
much higher than the white dwarf's orbital speed.  
The rotation of the white dwarf changes our viewing
angle, leading to large variations in velocity (up to a few thousand. km s$^{-1}$)
periodic on the white dwarf rotation period,  
which in co-rotating systems is the same as 
the orbital period $P_{\rm orb}$.
The brighter parts of the accretion column can also 
disappear over the limb of the white dwarf as
it rotates, causing the intensity of both the lines 
and the continuum to vary.
As with any binary system, eclipses also occur if 
the inclination is high enough. 
Typically, most of the eclipsed flux arises from the bright base
of the accretion column (or columns).

We have been observing CVs, mostly spectroscopically,
to characterize them and in particular to measure their
orbital periods when possible.
Here we present studies of 
\numberOfTargets CVs that are apparently AM Her stars.
Table~\ref{tab:star_info} lists the stars discussed here.

In Section \ref{sec:techniques}, we describe the 
instrumentation and techniques used for our
observations, reductions, and analysis.  Section 
\ref{sec:stars} gives detailed information on 
the individual objects.  Section \ref{sec:conclusions}
summarizes and draws attention to the results
we think are most interesting.

\begin{deluxetable}{llrrl}
\label{tab:star_info}
\tablewidth{0pt}
\tablecolumns{5}
\tablecaption{List of Objects}
\tablehead{
\colhead{Name} &
\colhead{$\alpha_{\rm ICRS}$} &
\colhead{$\delta_{\rm ICRS}$} &
\colhead{$G$} &
\colhead{$1/\pi_{\rm DR2}$} \\
\colhead{} &
\colhead{[h:m:s]} &
\colhead{[d:m:s]} &
\colhead{} &
\colhead{[pc]} \\
}
\startdata
Gaia18aot & 02:11:07.987 & +30:54:06.96 & 18.87 & $513 (+99, -72)$ \\
PT Per & 02:42:51.197 & +56:41:31.12 & 18.36 & $185 (+5,-4)$ \\
RX J0636.3+6554 & 06:36:22.915 & +65:54:14.77 & 18.86 & $438 (+52,-42)$ \\
CSS080228:081210+040352  & 08:12:10.239 & +04:03:51.43 & 18.71 & $1500(+800,-400)$ \\
SDSS J100516.61+694136.5 & 10:05:16.583 & +69:41:36.41 & 18.84 & $1640 (+850,-420)$ \\
SDSS J133309.20+143706.9 & 13:33:09.186 & +14:37:06.93 & 19.71 & \nodata \\
SDSS J134441.83+204408.3 & 13:44:41.834 & +20:44:08.61 & 18.29 & $591 (+74, -59)$ \\
Gaia18aya & 22:04:50.675 & +40:08:38.51 & 18.44 & $430 (+40, -34)$ \\
\enddata
\tablecomments{Positions, mean $G$ magnitudes, and distances
from the GAIA Data Release 2 (DR2; \citealt{GaiaPaper1,GaiaPaper2}).  
Positions are referred to the
ICRS (essentially the reference frame for J2000), and the catalog
epoch (for proper motion corrections) is 2015.  The distances and
their error bars are the inverse of the DR2 parallax $\pi_{\rm DR2}$,
and do not include any corrections for possible systematic errors.}
\end{deluxetable}

\section{Techniques}
\label{sec:techniques}

Nearly all of the data presented here are from MDM Observatory, on Kitt
Peak, Arizona.  Here we only summarize our observing protocols, 
data reduction, and analysis techniques, since they were
mostly similar to those described in previous papers (e.g. 
\citealt{halpern18,zoo}).

\subsection{Spectroscopy}

Most of our spectra are from the ``modspec''
spectrograph
\footnote{http://mdm.kpno.noao.edu/Manuals/ModSpec/modspec\_man.html},
usually mounted on the 2.4m Hiltner telescope, though 
occasionally on the 1.3m McGraw-Hill telescope.  A
600 line mm$^{-1}$ grating gave 2 \AA\ pixel$^{-1}$
with either of the two SITe CCD detectors ($2048^2$ or 
$1024^2$) we used. 
We reduced these data with IRAF software driven
by python scripts, but extracted the 2-dimensional
spectra to 1-dimensional spectra using our own implementation of 
the optimal extraction algorithm described by \citet{horne86}.
For wavelength calibration we derived a pixel-wavelength
relation from comparison lamps
taken in
twilight, and then adjusted the zero point using the 
[\ion{O}{1}] $\lambda$5577 airglow feature, since with this
instrument a linear shift accurately compensated the 
flexure
of the Cassegrain-mounted spectrograph
as the telescope moved. 

The most recent observations are from the Ohio State Multi-Object 
Spectrometer (OSMOS; \citealt{martini}) mounted on the 2.4 m,
using the blue grism and `inner' slit, which gave 
0.7 \AA\ pixel$^{-1}$ and $\sim 3$ \AA\ resolution.
While the reductions were generally similar to modspec, 
OSMOS required a more elaborate wavelength
calibration procedure since the 
pixel-to-wavelength scale was less stable.
To adjust the wavelength scale we either
measured airglow features (tabulated by
\citealt{osterbrock16}) or took short
Hg and Ne lamp exposures adjacent to our
science exposures.

We measured radial velocities, mostly of H$\alpha$, in the
individual exposures by by convolving the line 
profile with an antisymmetric function as described
by \citet{sy80}.  The choice of convolution function serves
to emphasize different parts of the line profile 
\citep{shafter83}.  For the most part we chose
the derivative of a Gaussian as the convolution function,
which provides a measure of the `overall' location of the line,
including the line core.

The emission lines of AM Her stars display complicated
profiles that change through their orbits.  We 
display these by creating two-dimensional images 
as follows.  In most cases, we start by rectifying
the spectra, that is, dividing them
by a smooth function fitted to the continuum.
Cosmic rays and other obvious artifacts are then
edited out by hand.  We compute the orbital phase
of each spectrum, divide the orbital cycle into 
100 phase bins and average together spectra 
that fall within a window of each phase point, 
using a weighting function that is a truncated
Gaussian in phase, centered on %
the phase point.
Finally, we stack the averaged spectra into a 
two-dimesional image, repeating a cycle to avoid
discontinuities.  The sources studied here are
rather faint, so some of our sources required exposure times of 720-900 s for adequate signal-to-noise; this resulted in some phase smearing.
Even so, most of the trailed 
emission line spectrograms show a rather sharp
component that is brightest as it swings from red
to blue.  This behavior is consistent with 
emission from the side of the companion star
irradiated by the X-ray and ultraviolet flux from the white dwarf
(see, e.g., \citealt{schwope97}, and for a very
early example, \citealt{tcmb78}).

For one of our targets, CSS080228:081210+040352, we also obtained four spectra with the Southern African Large Telescope \citep[SALT;][]{buck06}; these are described in Section \ref{subsec:css0812}.

\subsection{Photometry}

The MDM time-series
photometry is from the 1.3m telescope, mostly with an 
Andor IKON CCD frame-transfer CCD. Some of the 
photometric data were taken with a 1024$^2$-pixel
SITe CCD, cropped to a $256^2$ pixel subarray to 
reduce the CCD readout time.  The reduction script
performed aperture photometry on the program star, 
a comparison star, and several check stars in each
frame.

We also include some time-series photometry from
the 1-meter telescope at the South African Astronomical
Observatory (SAAO), taken using the Sutherland High-speed
Optical Camera (SHOC; \citealt{coppejans13}), using an Andor iXon 888 EM-CCD camera. 

The OSMOS spectroscopic target acquisition procedure
requires at least one direct exposure to place the 
slit on the target. We took these through a Sloan $g$ 
filter, and developed an automated program to infer the target's 
$g$ magnitude.  The script detects the stars in the image,
matches them to entries in the the PAN-STARRS
1 Data Release 2 catalog, performs aperture photometry,
establishes the offset between
the instrumental magnitude and the catalogued $g$,
and from this infers the $g$ magnitude of the target just before
the spectra were taken.  Because the offset between
instrumental and catalogued magnitude is common to 
the program and field stars, the procedure is 
differential.  Given adequate signal-to-noise,
it is accurate even in thin clouds and poor 
seeing. 

\section{The Individual Stars}
\label{sec:stars}

\subsection{Gaia18aot}
\label{subsec:gaia18aot}

This source was listed in the Gaia Transient Alerts
on 2018 March 07, with an alerting magnitude of 17.32.
Its Gaia light curve shows irregular fluctuations,
mostly between 18th and 19th magnitude, but sometimes
fainter than 20th.  The Catalina Real Time 
Survey Data Release 2 light curve is similar, but also
shows a brief flare on 2007 Nov.~02 that reaches 16.0.

\begin{figure}
\includegraphics[height=23 cm,trim = 1cm 4cm 0cm 5cm,clip=true]{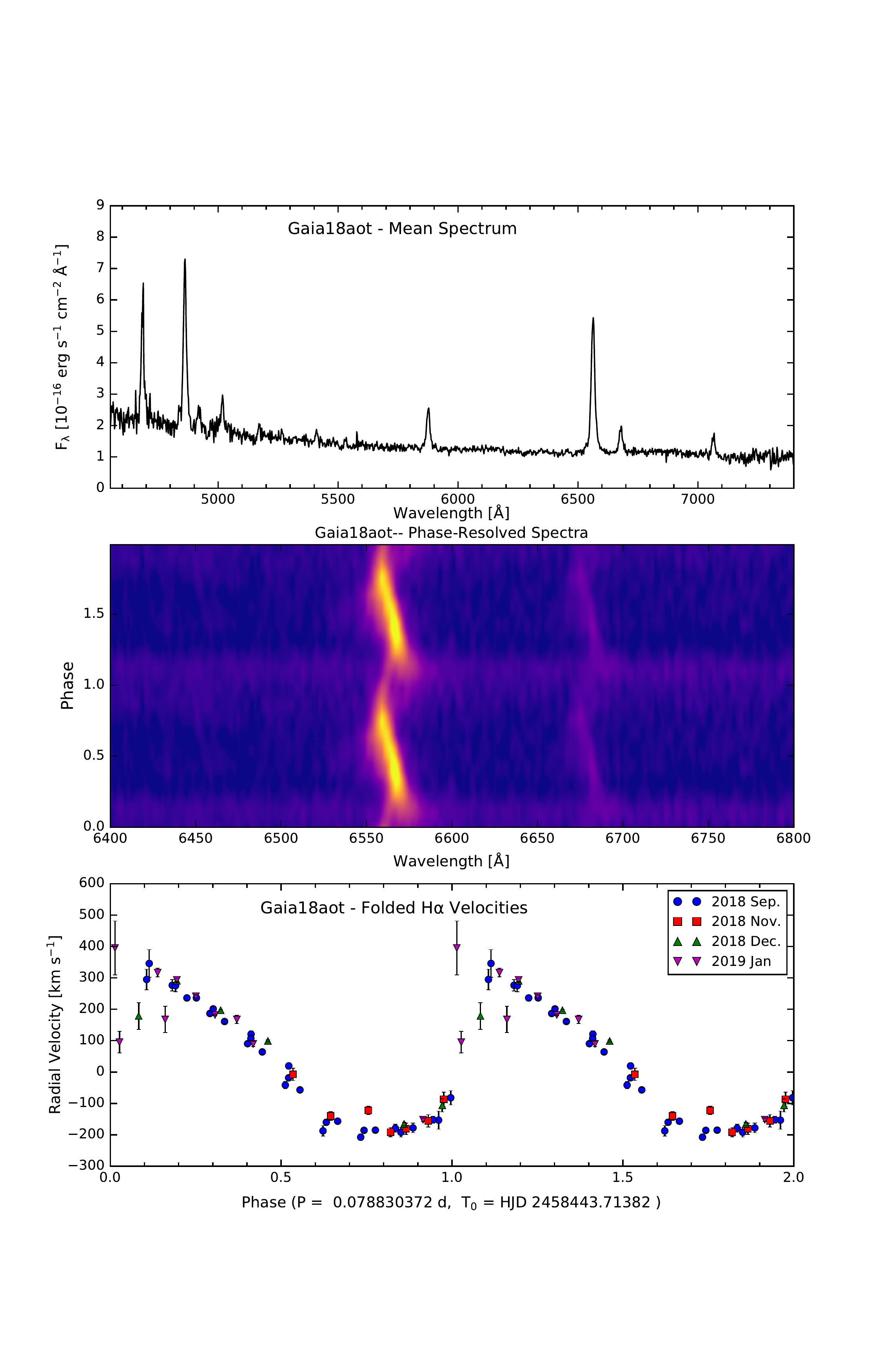}
\caption{Top:  Mean spectrum of Gaia18aot.  Middle: 
Phase resolved spectra of Gaia18aot in the region of 
H$\alpha$ and \ion{He}{1} $\lambda$6678.  Phase increases
from the bottom, and two cycles are shown for clarity. 
Lower: Velocities of H$\alpha$ folded on the
orbital ephemeris, with the best-fitting sinusoid
superposed.  
}
\label{fig:gaia18aotmontage}
\end{figure}

Most of our spectra are from 2018 September.
The mean spectrum (Fig.~\ref{fig:gaia18aotmontage}) shows
the strong emission in the Balmer, \ion{He}{1}, and 
\ion{He}{2} lines characteristic of magnetic CVs.  
The emission line velocities vary with $P \sim 114$ min,
in a non-sinusoidal pattern (Fig.~\ref{fig:gaia18aotmontage}.
We obtained some velocities in 
2018 November, 2018 December, and 2019 January. 
Combining these, we found $P = 0.078830(2)$ d, with 
no ambiguity in cycle count.
 
\begin{figure}
\label{fig:gaia18aotlightcurves}
\includegraphics[width=7 in]{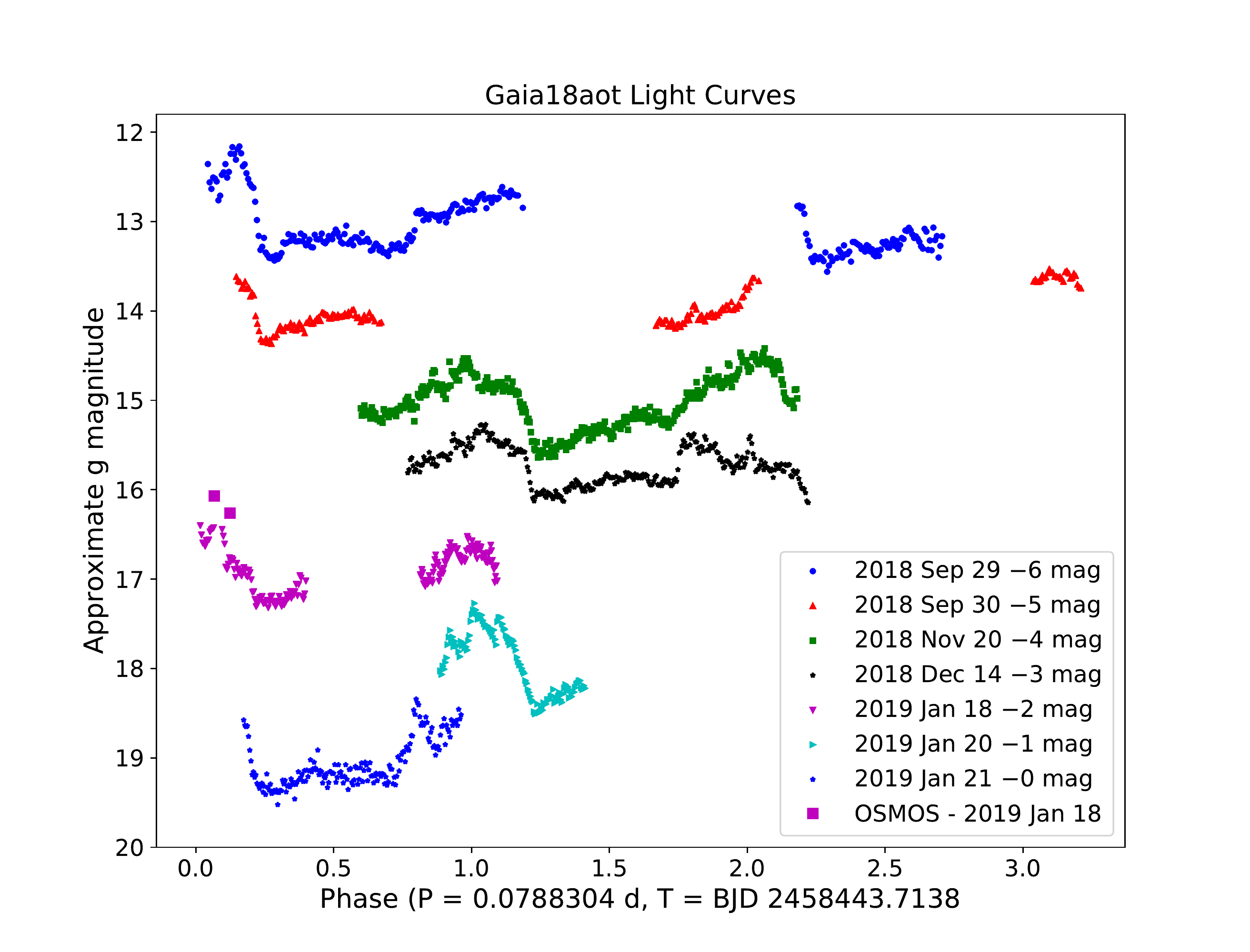}
\caption{Photometry of Gaia18aot from seven
nights spread over four observing runs.
The data are folded on the
spectroscopic orbital period, but with a new zero
in phase defined for each night so as to maintain 
time ordering through the night. 
Phase zero corresponds to blue-to-red crossing of the 
radial velocity. The earliest data are at the top,
and each night's data is offset downward by 1 magnitude,
as indicated in the legend.  The vertical scale is
correct for the lowermost trace.
The large squares, from OSMOS setup images, were
taken with a $g$ filter; the rest of the data are
essentially white-light, but adjusted to approximate
$g$ using the PAN-STARRS 1 magnitude of the comparison 
star.}
\label{fig:gaia18aot_tsphot}
\end{figure}

On the same observing runs, we obtained multiple orbits
of time-series photometry, which are summarized in 
Fig.~\ref{fig:gaia18aot_tsphot}.  The object remained
in a similar photometric state through all our runs, and
showed modulation at the orbital period,
most notably a rapid decrease in flux around phase 
0.2 in the radial-velocity ephemeris, and a more
gradual recovery around phase 0.7.  The small scatter
of the phase of the rapid decrease over multiple 
observing runs corroborates the already-secure choice
of cycle count.  


\subsection{PT Per}
\label{subsec:ptper}

\citet{watson16} review the history of this object, and
characterize it as a `relatively obscure and poorly studied
CV'.  They analyzed an 18-ksec serendipitous {\it XMM-Newton} 
observation of this source from 2011 July, 
and found deep minima in the X-ray
light curve, recurring with a 4900-s period.  They also 
obtained optical spectra from the William Herschel 
Telescope on three successive nights in 2015 April, at 
airmass $> 3$, and in evening twilight, which showed no
strong emission or absorption features, but did show 
weak, Zeeman-split absorption at H$\alpha$ and 
H$\beta$, consistent with a magnetic field of $\sim 25$ MG.  
They suggested that PT Per is a polar, and that their optical
spectra were taken in a low state.  Their observations
indicated a relatively small distance, perhaps as 
nearby as 90 pc, and indeed its Gaia DR2 distance 
of $1/\pi = 185(+5,-4)$ pc makes in the closest 
object studied here.

\begin{figure}
\includegraphics[height=23 cm,trim = 1cm 4cm 0cm 5cm,clip=true]{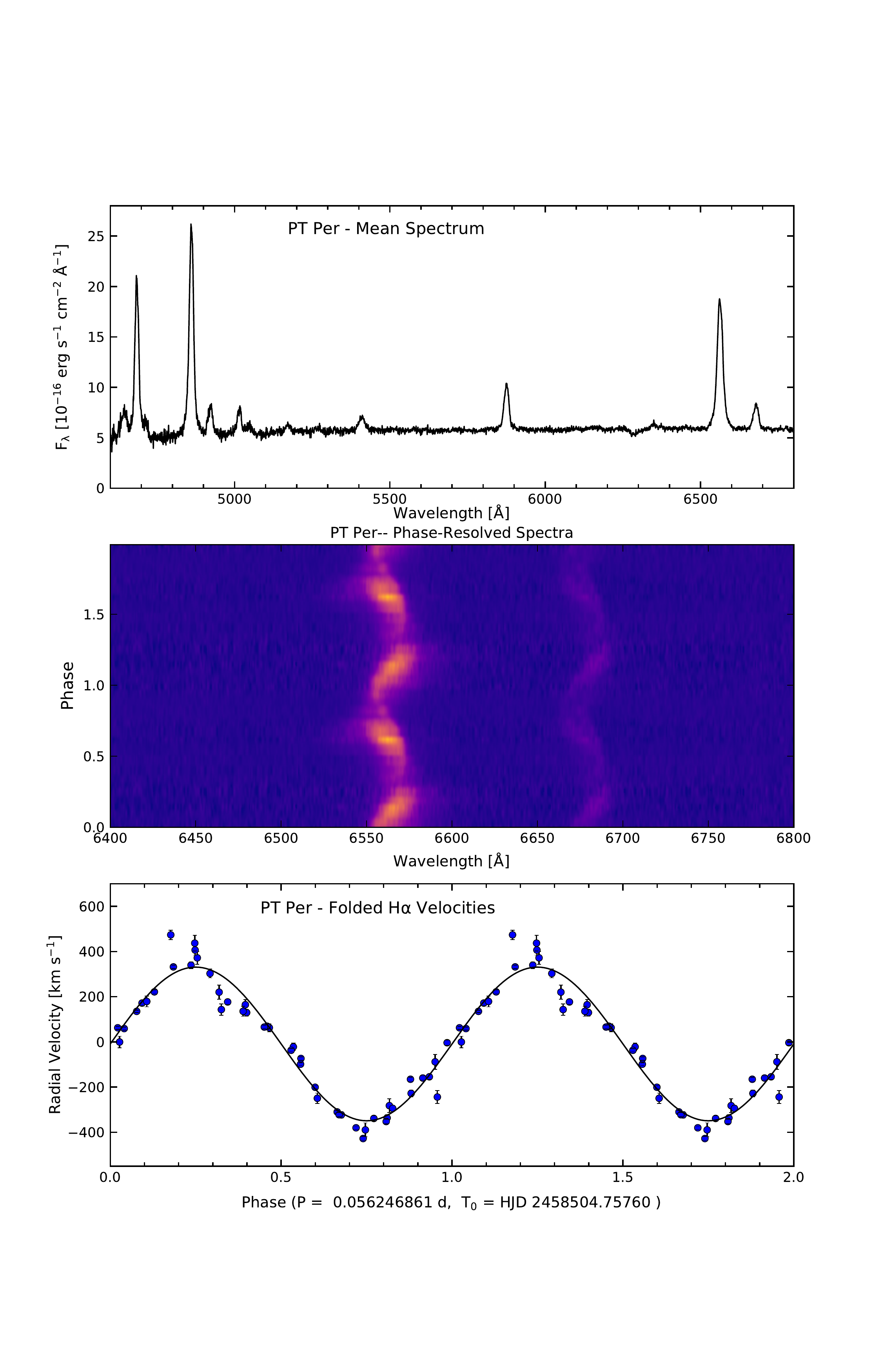}
\caption{Top:  Mean spectrum of PT Per.  Middle: 
Phase resolved spectra of PT Per in the region of 
H$\alpha$ and \ion{He}{1} $\lambda$6678.  Phase increases
from the bottom, and two cycles are shown for clarity. 
Lower: Velocities of H$\alpha$ folded on the
orbital ephemeris, with the best-fitting sinusoid
superposed.
}
\label{fig:ptpermontage}
\end{figure}

In 2019 January, we found PT Per in a much more active
state and obtained spectra with OSMOS on two successive nights. 
The spectrum (Fig.~\ref{fig:ptpermontage}, top) showed strong 
emission lines, in contrast to the \citet{watson16} spectra.
Large, rapid radial velocity shifts were immediately apparent; 
an analysis of the two nights' velocities gives a period 
$P = 81.00(4)$ min, with no cycle-count ambiguity, in 
reasonable agreement with the 81.7(4) min
period found by \citet{watson16}.  The phase-resolved spectra
in the middle panel of Fig.~\ref{fig:ptpermontage} shows the 
large velocity shifts, as well as the asymmetric line wings 
characteristic of AM Her stars. Large blue-shifted velocity excursions are also seen, near phase $\sim$0.7, typical of polars. The H$\alpha$ radial velocities 
(lower panel) are modulated almost sinusoidally. 
The velocity half-amplitude, $K = 340 \pm 14$ km
s$^{-1}$, is much too large to be plausibly orbital, so the
infall velocity of the accretion column causes most
of the velocity shift.

The high-state data amply confirm that PT Per is an AM Her
star, as suggested by \citet{watson16}.  They note that their
optical data were taken in a remarkably low state, with no
clear emission lines, whereas most AM Her stars continue to show
some emission lines even in very low states.  



\begin{figure}
\includegraphics[width=7 in]{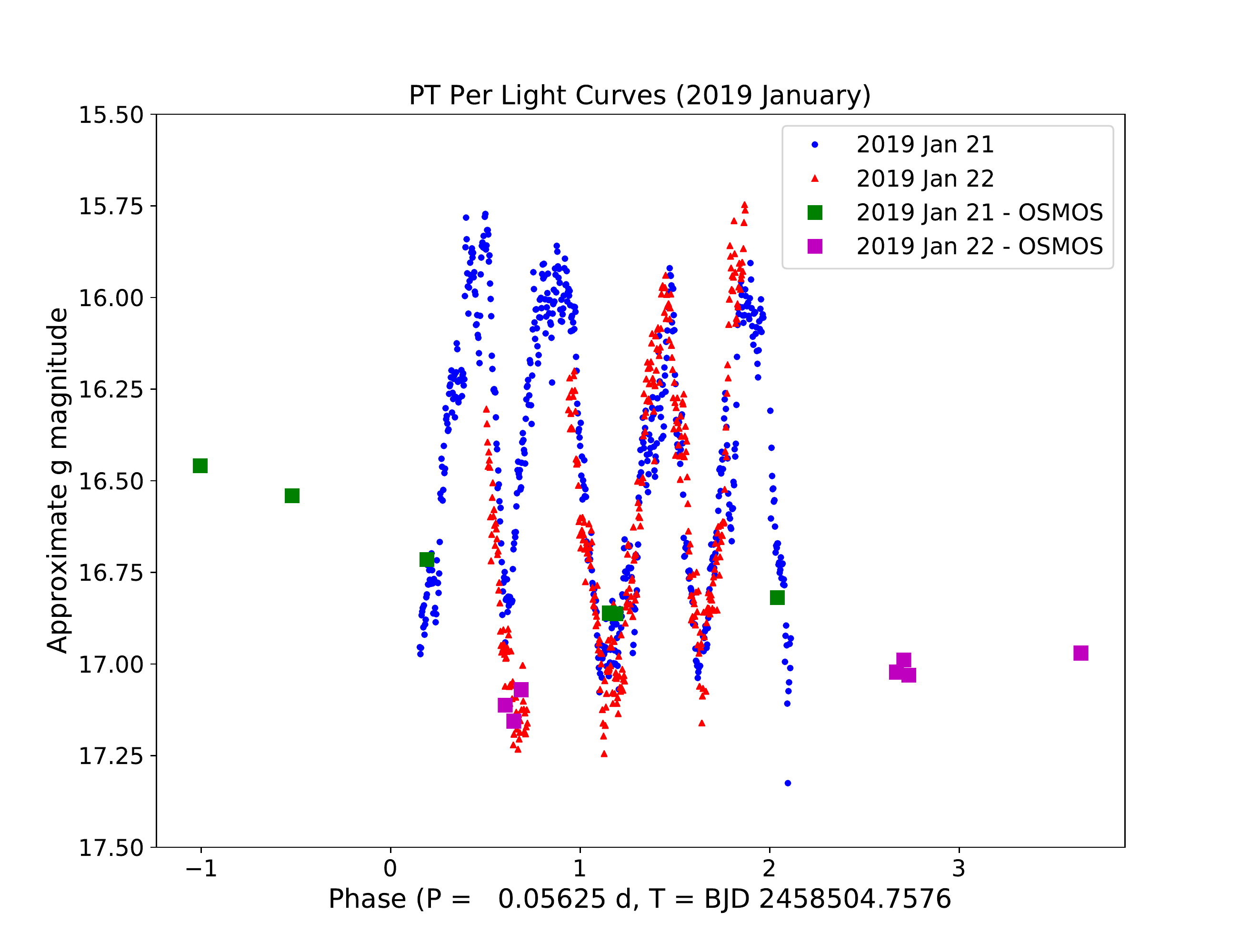}
\caption{Photometry of PT Per from two
successive nights.  The phase is treated as in 
Fig.~2.  Phase zero corresponds to blue-to-red crossing of the 
radial velocity. No vertical offset is applied.
The large squares, from OSMOS setup images, were
taken with a $g$ filter; the time-series points 
were taken almost unfiltered and were adjusted to 
agree with the calibrated OSMOS 
magnitudes obtained simultaneously.
}
\label{fig:ptper_tsphot}
\end{figure}

On 2019 Jan 21 and 22, we used the 1.3 m telescope and Andor
camera to obtain the 
time-series photometry shown in Fig.~\ref{fig:ptper_tsphot}.
The light curve shows with two maxima per orbit, along with 
some flickering.  There is no sign of an eclipse,
so the correspondence between the orbital phase plotted
and the locations of the stars in their orbits is 
not constrained.  The double-humped light curve
indicates that accretion likely occurs onto two poles.


\subsection{RX J0636.3+6554} 
\label{subsec:rx0636}

\citet{appenzeller98} discovered this star as the optical counterpart
of a ROSAT X-ray source; they noted it was blue, variable on a timescale
of hours, and that one of their spectra showed broad H$\alpha$
emission at rest velocity. It was listed in the \citet{downes01} catalog, but apparently
no follow-up studies have appeared.  The CRTS Data Release 2 light 
curve \citep{crts} shows short term variation of about 1 magnitude
superposed on a gradual decline from $\sim$ 17.6 mag in 2006 to 
about 20.0 mag in 2013.  

\begin{figure}
\includegraphics[height=23 cm,trim = 1cm 4cm 0cm 5cm,clip=true]{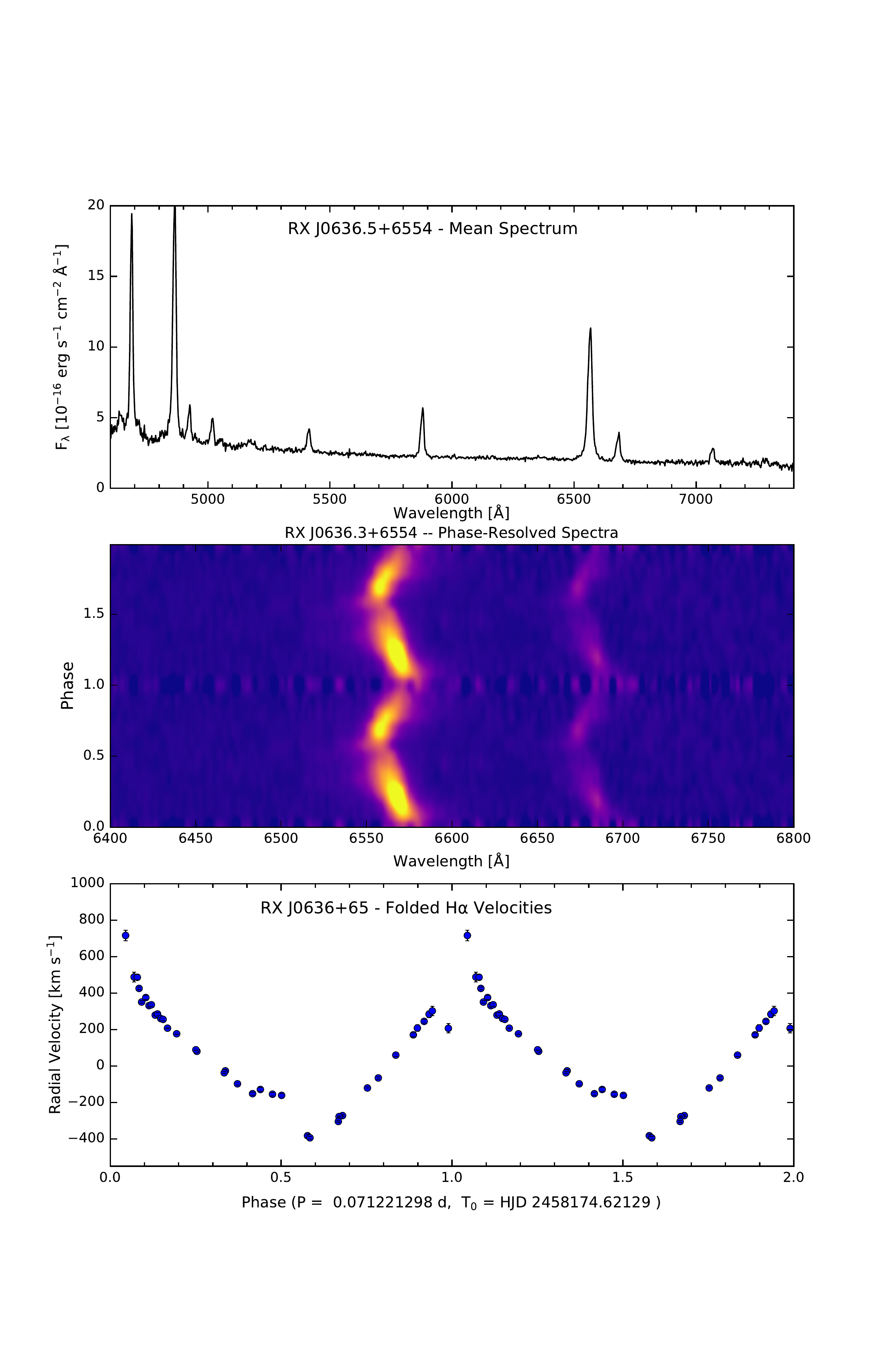}
\caption{Top:  Mean spectrum of RX J0636.3+6554.  Middle: 
Phase resolved spectra of RX0636.3+6554 in the region of 
H$\alpha$ and \ion{He}{1} $\lambda$6678.  Phase increases
from the bottom, and two cycles are shown for clarity.  The
horizontal discontinuities are caused by incomplete phase
coverage, and the noise near phase 0 is largely caused by the
eclipse.  Lower: Velocities of H$\alpha$ folded on the
orbital ephemeris, showing large velocity swings and a
non-sinusoidal modulation.}
\label{fig:rx0636montage}
\end{figure}

We took spectra of this star in 2018 February.  The mean 
spectrum (Fig.~\ref{fig:rx0636montage}, top panel) shows strong emission 
lines on a blue continuum, with \ion{He}{2} $\lambda 4686$ nearly
as strong as H$\beta$.  Passing the fluxed spectrum through the
$V$ response function tabulated by \citet{bessell90} gives $V \sim 17.9$,
so we caught the system in a relatively bright state.  The 
emission lines immediately showed large velocity swings on a period
just over 100 min (Fig.~\ref{fig:rx0636montage}, middle and 
lower panels).

\begin{figure}
\includegraphics[width=7 in]{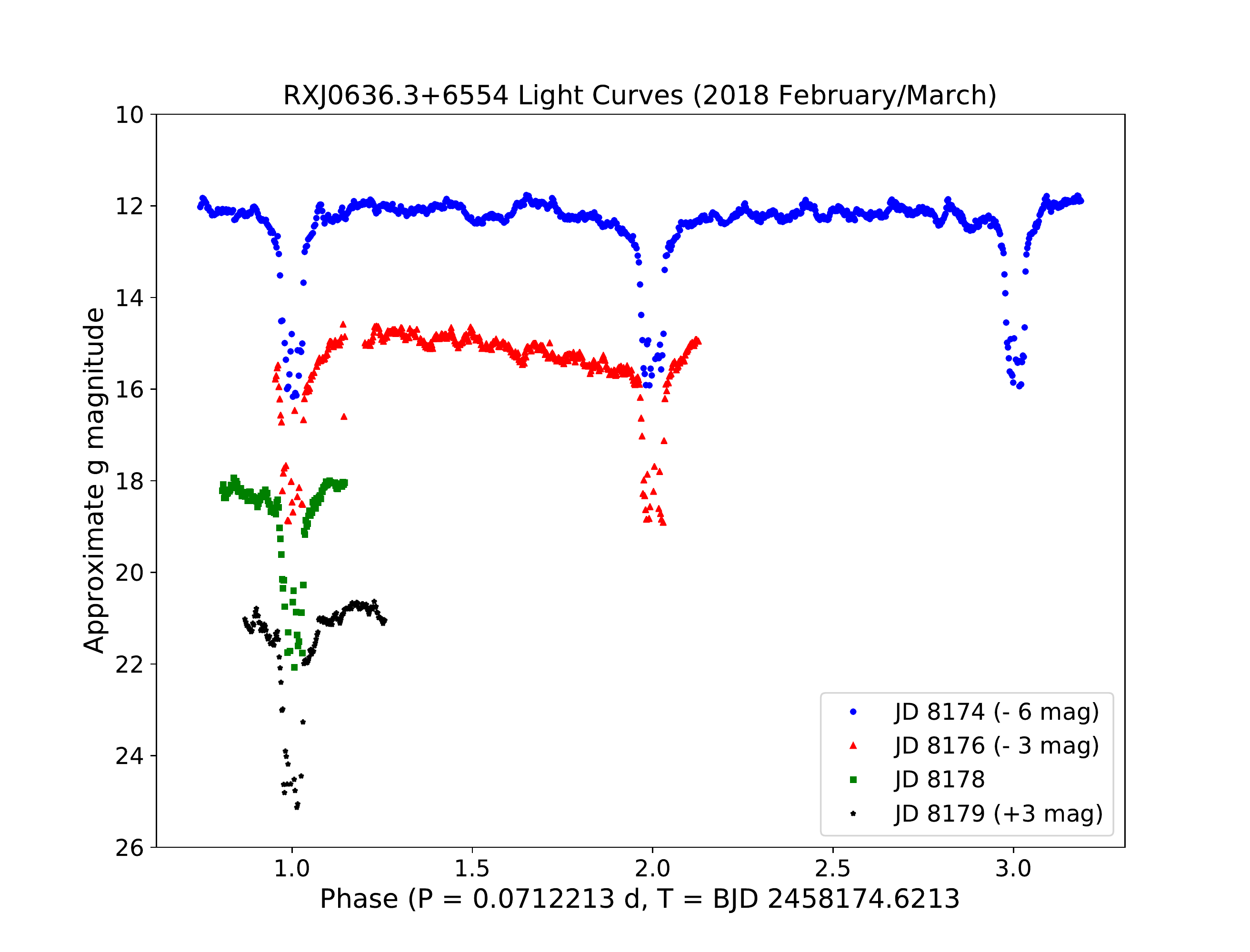}
\caption{Differential photometry of RX J0636.3+6554 on
four nights during the brighter state, folded on the best-fit linear ephemeris (eqn. \ref{eqn:rx0636ephem}).
The curves are offset vertically by 3 mag.  The data were taken with a 
UV cutoff filter only; 
the $g$ magnitude of the reference star has been added to convert
to a rough $g$ magnitude.}
\label{fig:rx0636eclipsefig}
\end{figure}

The star disappeared from time to time during the spectroscopy,
so we obtained time-series photometry on the 
same observing run (see Fig.\ref{fig:rx0636eclipsefig}).  
This showed eclipses $\sim 2$ mag deep and lasting $\sim$6 min 
on a $\sim 103$ min period, as well as out-of-eclipse
flickering.  Over the last two years
we have observed 23 eclipses, including two that were 
generously observed by Karolina B\k{a}kowska in 2018 April.
Table \ref{tab:rx0636eclipses} gives the times of eclipse
center, along with the cycle count and residuals from the 
best linear ephemeris, which is 
\begin{equation}
\label{eqn:rx0636ephem}
\hbox{BJD mid-eclipse} = 2458174.62129(5) + 0.071221298(9). 
\end{equation}
where $E$ is an integer eclipse number and the time base is UTC.
Nearly all our time-series photometry 
was relative to a star $\sim 81$ arcsec from the program
object in position angle $\sim 63^{\circ}$, for which 
Gaia DR2 lists $\alpha$ = 6:36:34.75 and $\delta = $+65:54:51.7.
The PAN-STARRS data release 2 gives gives $g_{\rm PSF} = 17.07$
for this star, which we added to our differential magnitudes.
During our 2018 February and April observations
, the out-of-eclipse
magnitude averaged $g \sim 18.1$, while for all our other
time-series photometry it was much fainter at $\sim 20.1$.

\begin{deluxetable}{rrrr}
\label{tab:rx0636eclipses}
\tablecolumns{4}
\tablecaption{RX J0636.3+6554 Eclipse Centers}
\tablehead{
\colhead{$E$} &
\colhead{Time} &
\colhead{$O-C$} &
\colhead{Date} \\
\colhead{} &
\colhead{} &
\colhead{[s]} &
\colhead{} \\
}
\startdata
$     0$ &    8174.62125 & $  -3$\phm{xx} & 2018-02-25 \\
$     1$ &    8174.69239 & $ -11$\phm{xx} & 2018-02-25 \\
$     2$ &    8174.76399 & $  22$\phm{xx} & 2018-02-25 \\
$    28$ &    8176.61553 & $   4$\phm{xx} & 2018-02-27 \\
$    29$ &    8176.68680 & $   8$\phm{xx} & 2018-02-27 \\
$    56$ &    8178.60964 & $  -3$\phm{xx} & 2018-03-01 \\
$    70$ &    8179.60681 & $   2$\phm{xx} & 2018-03-02 \\
$   647$ &    8220.70147 & $  -0$\phm{xx} & 2018-04-12 \\
$   688$ &    8223.62175 & $  18$\phm{xx} & 2018-04-15 \\
$  3710$ &    8438.85216 & $ -13$\phm{xx} & 2018-11-16 \\
$  3724$ &    8439.84923 & $ -15$\phm{xx} & 2018-11-17 \\
$  3753$ &    8441.91472 & $  -9$\phm{xx} & 2018-11-19 \\
$  3754$ &    8441.98586 & $ -15$\phm{xx} & 2018-11-19 \\
$  4088$ &    8465.77391 & $  -4$\phm{xx} & 2018-12-13 \\
$  7840$ &    8732.99620 & $  -6$\phm{xx} & 2019-09-06 \\
$  8485$ &    8778.93416 & $  14$\phm{xx} & 2019-10-22 \\
$  8486$ &    8779.00504 & $ -16$\phm{xx} & 2019-10-22 \\
$  8499$ &    8779.93109 & $  -1$\phm{xx} & 2019-10-23 \\
$  8512$ &    8780.85697 & $  -1$\phm{xx} & 2019-10-24 \\
$  9299$ &    8836.90813 & $  -0$\phm{xx} & 2019-12-19 \\
$  9662$ &    8862.76148 & $   1$\phm{xx} & 2020-01-14 \\
$  9663$ &    8862.83294 & $  22$\phm{xx} & 2020-01-14 \\
$  9718$ &    8866.74993 & $   6$\phm{xx} & 2020-01-18 \\
\enddata
\tablecomments{Observed times of mid-eclipse.  The first
column gives the eclipse number $E$, and the second the barycentric
julian date minus 2 450 000., on the UTC system.  The penultimate
column gives the residual compared to the best-fit linear ephemeris
(eqn. \ref{eqn:rx0636ephem}), and the last the calendar date in UT.}
\end{deluxetable}

\begin{figure}
\includegraphics[width=6 in]{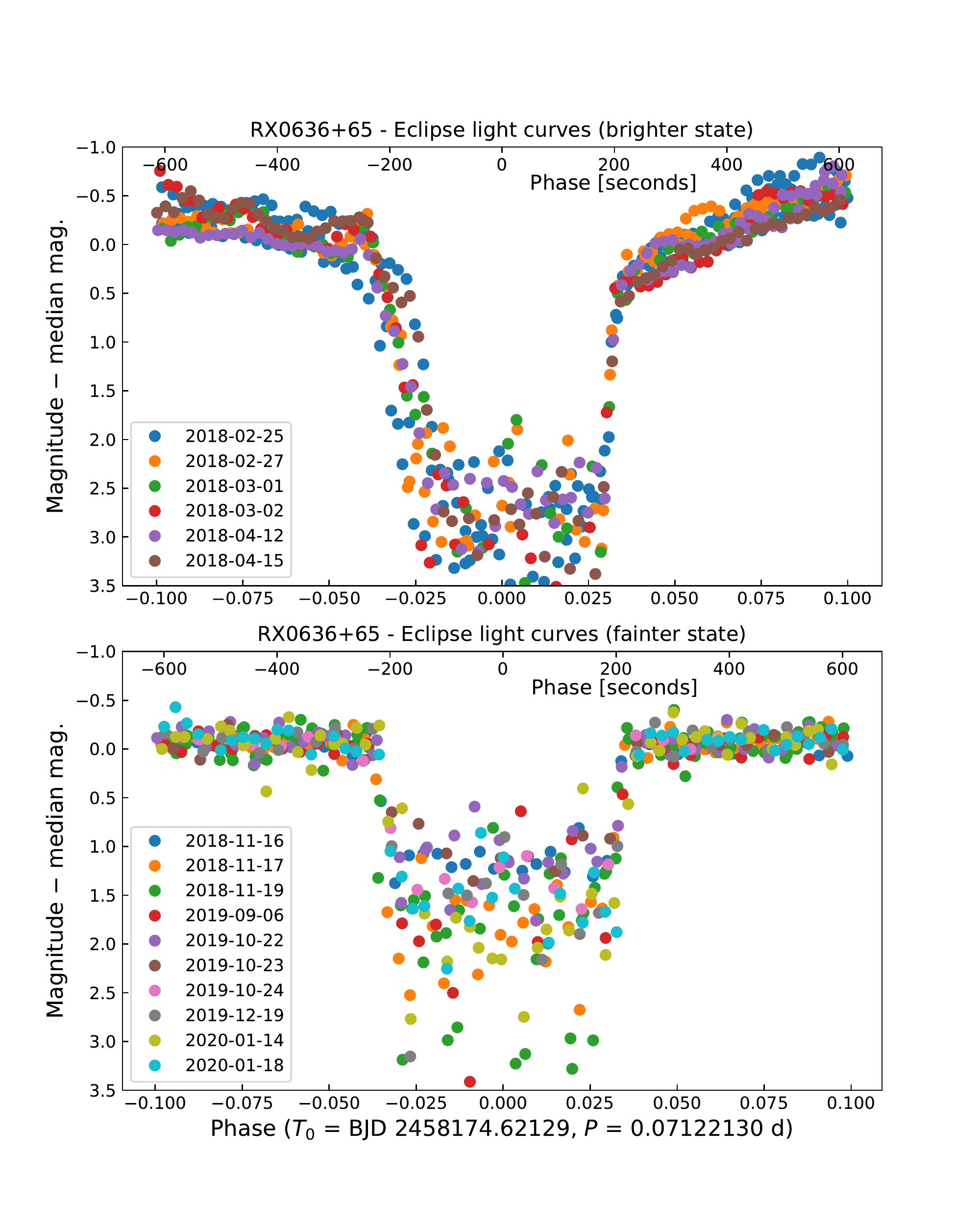}
\caption{Differential photometry of RX J0636.5+6554 in the vicinity
of the eclipse.  Measurements from many different nights
are superposed.  The measurements from each eclipse are 
adjusted so that their median is zero.  The horizontal
axis is from eqn.~\ref{eqn:rx0636ephem}, and 
the upper axis in each panel gives the offset from 
mid-eclipse in seconds. The upper panel shows data 
from the brighter state ($g \sim 18.1$) while the 
lower panel is from the fainter state ($g \sim 20.1$).
}
\label{fig:rx0636eclipsedetail}
\end{figure}

Fig.~\ref{fig:rx0636eclipsedetail}
shows the brighter- and fainter-state eclipse light
curves in greater detail. In both the brighter and
fainter states, the egress is sharply defined and occurs
at a very consistent phase.
Our exposures (typically 
20 sec) do not resolve the sharp rise in egress.
In the fainter
state, the ingress is also very consistent, but in the 
brighter state there is significant dispersion in the 
ingress phase.  This suggests that in the bright state,
a significant source of light lags behind the trailing
side of the white dwarf; a natural candidate for this
would be an accretion stream that fades away during
the faint state. In some light curves, the ingress starts slightly earlier than in others, suggesting an extra source of obscuration, which might be the outermost parts of the accretion stream.


We also note that in the bright-state egress, 
following the initial rapid rise, the object consistently undergoes a
slower, steady brightening. This may be explained by
the gradual uncovering of the inner part of the
magnetically threaded accretion stream.

The eclipse in the fainter state appears to be that of
the white dwarf alone, and we conservatively estimate the 
full width as $412 \pm 8$ seconds.  If the secondary 
star fills its Roche lobe, this implies a minimum
$q = M_2 / M_{\rm WD}$ of 0.11 to 0.13
for an edge-on orbit \citep{chanan76}.  In non-magnetic
systems, the relationship between $P_{\rm orb}$ and 
$q$ has been calibrated at short periods (see e.g.
\citealt{pattmurmurs}); dwarf novae at this period
have $q < 0.2$, which if applicable here
implies an upper limit for $i$ of $\sim 83^{\circ}$.
Assuming an implausibly large $q = 1$ gives $i \sim 73^{\circ}$.

Taking $90^{\circ} \le i < 83^{\circ}$ constrains
the dynamically important quantity $\sin^3 i$ to 
an accuracy of better than 3 per cent, so if we did
have a reliable measurement of the secondary's
velocity amplitude $K_2$, we could in principle 
determine $M_{\rm WD}$ quite accurately. Although the
secondary is likely to be extremely faint, similar
systems often have a strong, narrow component in their
emission line profiles that arises on the side
of the secondary facing the white dwarf.  Our 
spectroscopy shows a hint of this, but at our 
spectral resolution it is not cleanly defined, 
so we are unable to draw any useful conclusions.  

\subsection{CSS080228:081210+040352}
\label{subsec:css0812}
This object (abbreviated CSS0812+04) was detected in 2008 by the 
Catalina sky survey at a magnitude of 
18.8, and listed as an eclipsing CV with a Sloan Digital
Sky Survey (SDSS) magnitude of 
22.4.  However, the PAN-STARRS 1 survey consistently detects it
with $18.0 < g < 19.1$, suggesting that SDSS caught it in a state
of low mass transfer. We selected this object in 2014 for further 
photometric and spectroscopic studies after it was identified by one of us, MM, 
as a candidate polar based on its long-term CRTS light curve; this is part of a study 
to identify candidate polars based on long-term photometric behavior. Independently, 
\citet{oliveira20} obtained a survey
spectrum and classified it as a magnetic system.

\begin{figure}
\includegraphics[height=23 cm,trim = 1cm 4cm 0cm 5cm,clip=true]{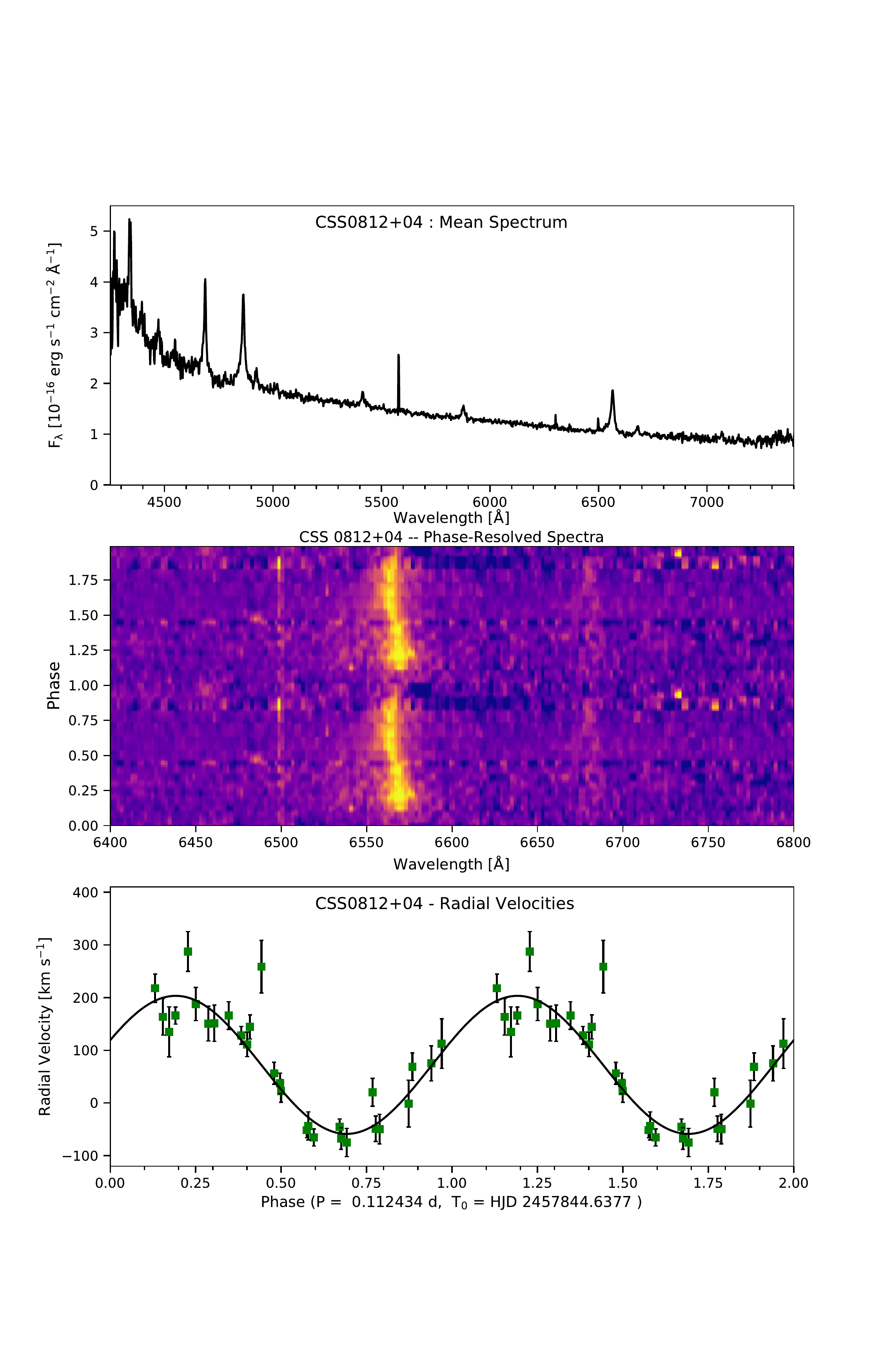}
\caption{Top:  Mean spectrum of CSS0812+04.  Middle: 
Phase resolved spectra of CSS0812+04 in the region of 
H$\alpha$ and \ion{He}{1} $\lambda$6678.  Phase increases
from the bottom, and two cycles are shown for clarity.  Lower: 
Velocities of H$\alpha$ folded on the ephemeris of the
sharp dip, with the best-fitting sinusoid overplotted.}
\label{fig:css0812+04specplot}
\end{figure}

\begin{figure}
\centering
\includegraphics[width=7in]
{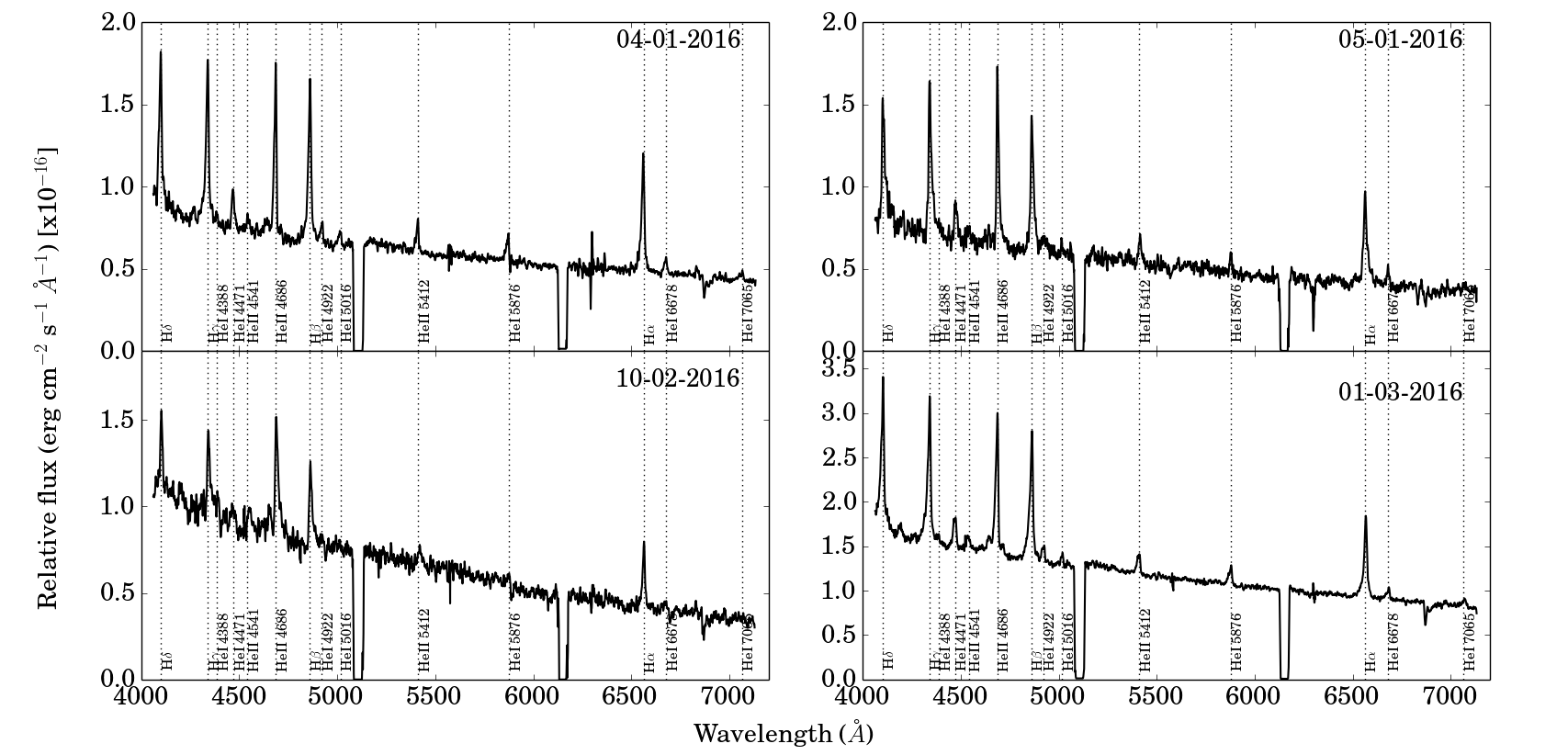}
\caption{SALT RSS spectra of CSS0812+04. The two gaps in the spectra at $\sim$5100\AA\ and ~6150\AA\ are due to gaps in the CCD detector mosaic.}
\label{fig:css0812+04SALTspec}
\end{figure}

Our mean spectrum from 2017 March (Fig.~\ref{fig:css0812+04specplot}) 
shows strong
\ion{He}{2} $\lambda4686$ emission, and also emission at 
$\lambda$ 5411, which are typical of magnetic CVs.
%
%
%
%
Four additional spectra were taken with the Robert Stobie Spectrograph (RSS) \citep[RSS;][]{burg03,kobu03,smit06} on the Southern African Large Telescope \citep[SALT;][]{buck06} on 4 \& 5 January, 10 February and 1 March 2016. The RSS was used in long-slit spectroscopy mode with a slit width 1.5''. The PG900 VPH grating was used, set to an incidence angle of 14.75$^{\circ}$, giving a spectral coverage of 4060-7100\,$\AA{}$ and a mean resolution of 5.7 \AA. On each night, 2$\times$690\,s exposures were taken. The wavelength calibration was done using Ar lamp exposures taken immediately following the observations and relative flux calibration was achieved using standard stars LTT\,377 and LTT\,4364, depending the night of observation. The SALT spectra are shown in Fig.~\ref{fig:css0812+04SALTspec}.

Radial velocities taken over 
two nights gave $P = 162.0 \pm 0.3$ min.  The velocity half-amplitude 
$K = 131 \pm 14$ km s$^{-1}$
is much smaller than expected for an AM Her star.

\begin{figure}
\includegraphics[width=7 in]{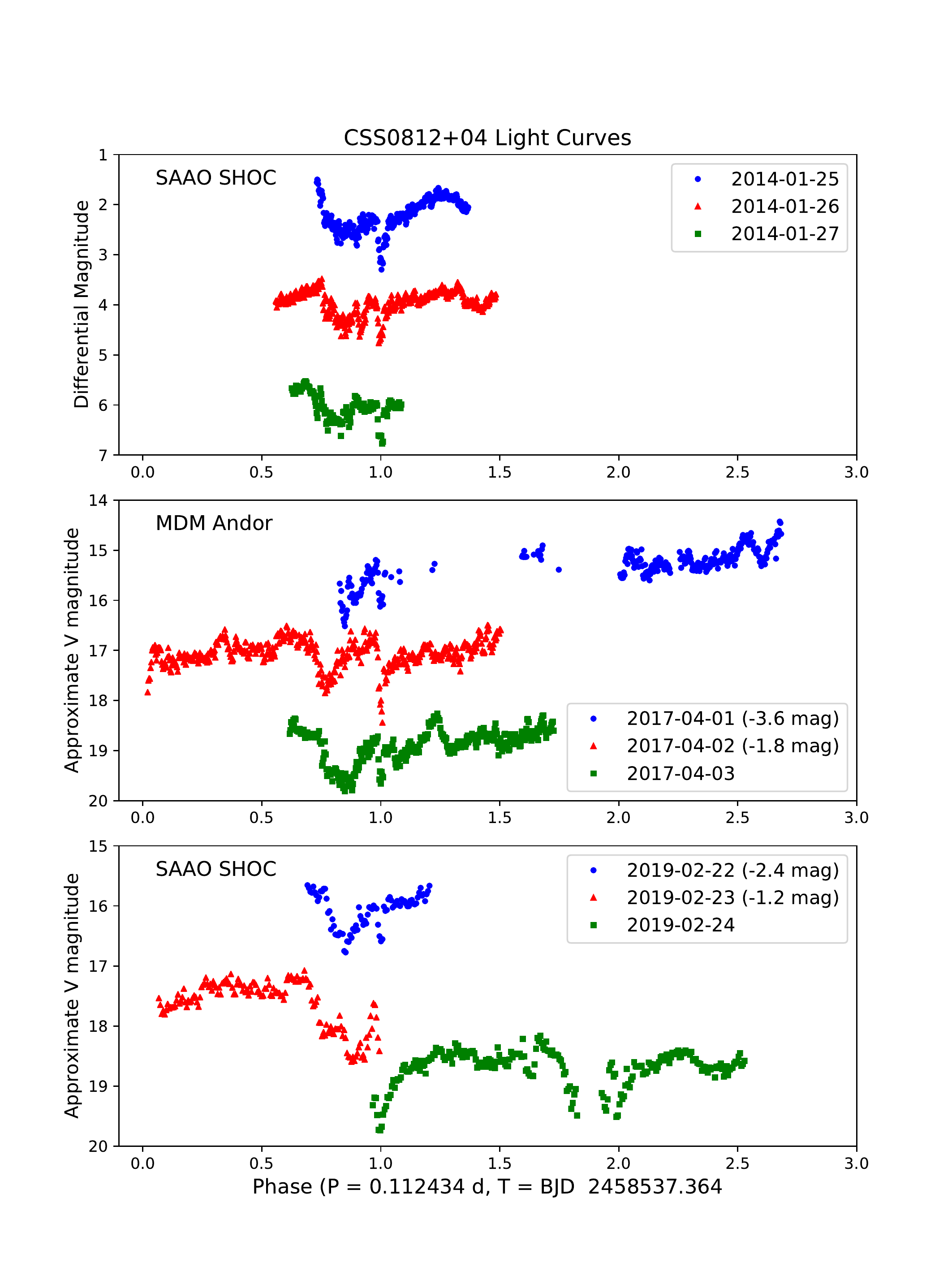}
\caption{Differential photometry of CSS0812+04, from three
different observing runs, aligned on the sharp dip
ephemeris (eqn. \ref{eqn:css0812ephem}).}
\label{fig:css0812+04lightcurves}
\end{figure}

Fig.~\ref{fig:css0812+04lightcurves} shows a sampling of our
time-series photometry.  Many of our light curves show a
very short dip, resembling a partial eclipse.  This feature
appeared insignificant until we found the spectroscopic
period, which made it evident that 
dips on successive nights were separated by integer
multiples of the orbital period.  We were able to 
connect dips found in 2014, 2017, and 2019 with a 
unique ephemeris, 
\begin{equation}
\label{eqn:css0812ephem}
\hbox{BJD sharp dip} = 2457844.6379(2) + 0.11241902(2) E,
\end{equation}
which we take to be orbital; Table \ref{tab:css0812dips}
gives the observed dip times and their assigned cycle numbers.
In many light curves, a broader minimum occurs
shortly before the dip.  The phase of this decline 
is consistent to better than $\sim$ 0.05 cycle, which
corroborates our choice of period.  

\begin{figure}
\includegraphics[width=7 in]{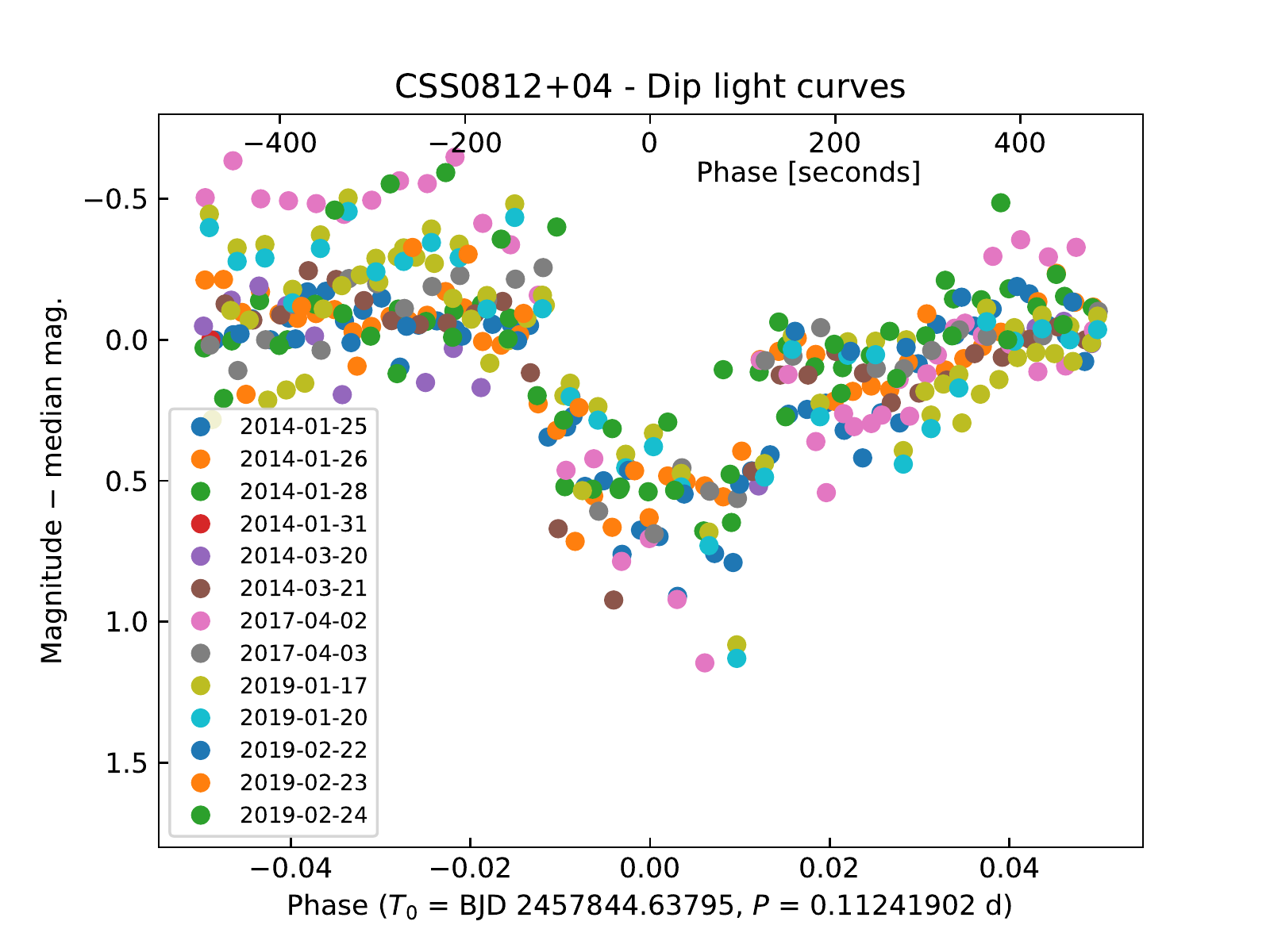}
\caption{A magnified view of the sharp dip, 
showing differential magnitudes from 13 different nights.
The data were aligned by subtracting the median value
of each night's magnitudes.
The lower scale is in cycles, and the upper
in seconds (see eqn. \ref{eqn:css0812ephem}).}
\label{fig:dipfig}
\end{figure}

Fig.~\ref{fig:dipfig} is a close-up view of the dip,
with data from 13 nights plotted.  The dip appears to 
be stable in phase, about 250 seconds wide, and typically
about 0.6 mag deep.  Its consistency suggests it is 
caused by a grazing
eclipse of the bright accretion column by the secondary
star.  A  compact accretion column disappearing 
momentarily over the 
limb of a rotating white dwarf might, in principle, mimic 
the dip's appearance, but such events tend to 
have more gradual ingresses and egresses, and not
to be as consistent.

\begin{deluxetable}{rrrr}
\label{tab:css0812dips}
\tablecolumns{4}
\tablecaption{CSS0812+04 Dip Timess}
\tablehead{
\colhead{$E$} &
\colhead{ Time } &
\colhead{ $O-C$ } &
\colhead{Date} \\
\colhead{} &
\colhead{} &
\colhead{[s]} &
\colhead{} \\
}
\startdata
$-10329$ & 6683.46241 & $  41$\phm{xx} & 2014-01-25 \\
$-10320$ & 6684.47391 & $  18$\phm{xx} & 2014-01-26 \\
$-10310$ & 6685.59779 & $  -9$\phm{xx} & 2014-01-28 \\
$ -9841$ & 6738.32191 & $ -43$\phm{xx} & 2014-03-21 \\
$     0$ & 7844.63770 & $ -21$\phm{xx} & 2017-04-01 \\
$    10$ & 7845.76210 & $  -3$\phm{xx} & 2017-04-02 \\
$    18$ & 7846.66150 & $   1$\phm{xx} & 2017-04-03 \\
$  5864$ & 8503.86368 & $  54$\phm{xx} & 2019-01-20 \\
$  6171$ & 8538.37549 & $ -18$\phm{xx} & 2019-02-23 \\
$  6179$ & 8539.27510 & $   5$\phm{xx} & 2019-02-24 \\
$  6180$ & 8539.38719 & $ -24$\phm{xx} & 2019-02-24 \\
\enddata
\tablecomments{Observed times of the sharp dip.  The first
column gives the dip number $E$, and the second the barycentric
julian date minus 2 450 000., on the UTC system.  The penultimate
column lists the residual compared to the best-fit linear ephemeris
(eqn. \ref{eqn:css0812ephem}), and the last gives the UT date.}
\end{deluxetable}
                       
\subsection{SDSS J100516.61+694136.5}
\label{subsec:sdss1005}

\citet{wils10} discovered this object (hereafter SDSS1005+69) 
by mining data from SDSS, Galaxy Evolution Explorer (GALEX), 
and various astrometric catalogs
for dwarf nova candidates.  They noted 
strong emission lines, including \ion{He}{2} $\lambda$4686 
comparable to H$\beta$, in the SDSS spectrum,
and suggested that it is a magnetic CV, varying from  
17.9 through 21.2 mag.

\begin{figure}
\includegraphics[height=23 cm,trim = 1cm 4cm 0cm 5cm,clip=true]{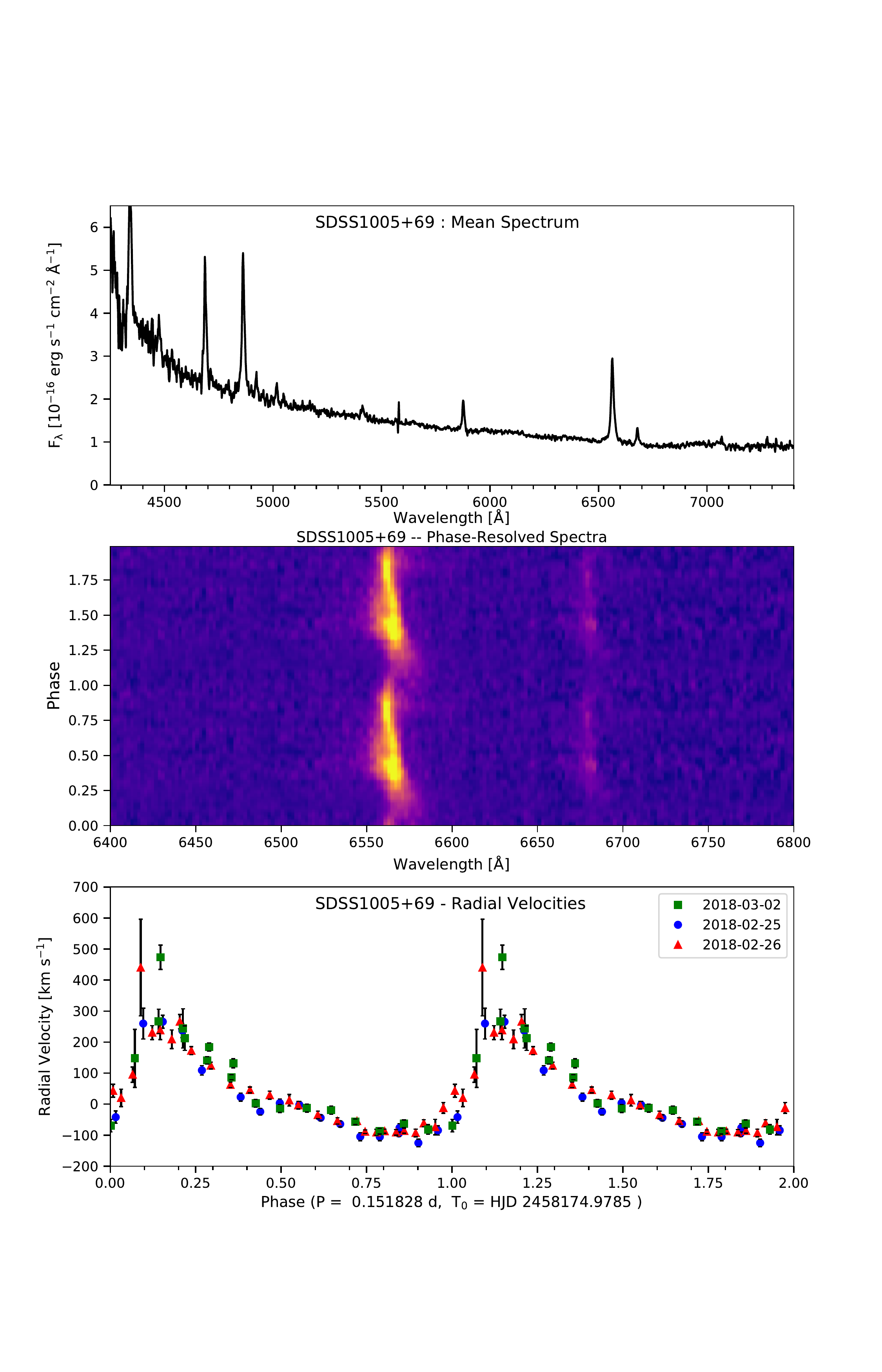}
\caption{Top:  Mean spectrum of SDSS1005+69 from 2018 February and March.  Middle: 
Phase resolved spectra of SDSS1005+69 in the region of 
H$\alpha$ and \ion{He}{1} $\lambda$6678.  Phase increases
from the bottom, and two cycles are shown for clarity.  Lower: 
Velocities of H$\alpha$, folded on the apparent orbital
period. 
}
\label{fig:sdss1005montage}
\end{figure}

\begin{figure}
\includegraphics[width=6.5 in]{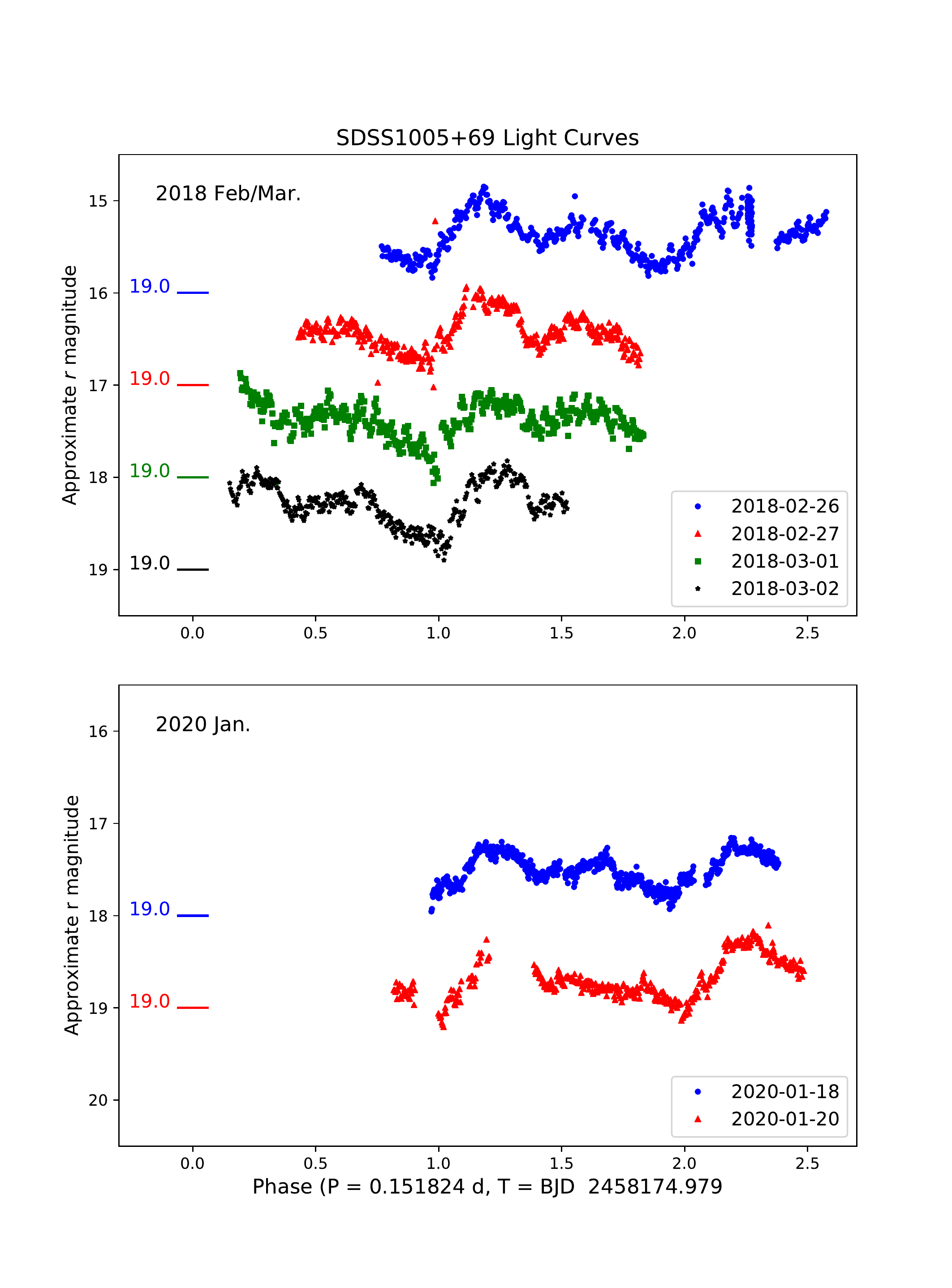}
\caption{Light curves of SDSS1005+69 from two observing runs.  The differential
magnitudes have been adjusted by the comparison star's $r$ magnitude.
The lowermost trace in each panel is plotted without a vertical offset,
and successive traces offset upward by 1.0 mag.  The period is not
known to the accuracy shown, but is 
fine-tuned to align the light curves from the two observing runs. 
}
\label{fig:sdss1005lightcurves}
\end{figure}

We obtained single spectra in 2012 January and 2015 April,
but did not find the system bright enough to study.
We enjoyed better luck in 2018 February and March and obtained spectra
on three nights.  The top panel of Fig.~\ref{fig:sdss1005montage} shows
the mean spectrum, which includes the \ion{He}{2} emission
characteristic of magnetic CVs.  The H$\alpha$ emission
line velocities are strongly modulated at an unambiguous
period of 218.6(4) min; the modulation is non-sinusoidal
with a rapid rise and a more gradual decline in each
cycle.  The phase used in the lower panel of 
Fig.~\ref{fig:sdss1005montage} is based on a sinusoidal
fit to the velocity data, and is essentially arbitrary.
Using the 1.3m telescope, we obtained time-series photometry
contemporaneous with our 2018 spectroscopy, and also on 
two nights in 2020 January 
(see Fig.~\ref{fig:sdss1005lightcurves}).  The comparison star
was at $\alpha = 10^{\rm h} 05^{\rm m} 14^{\rm s}.34, 
\delta = +69^\circ 43' 23''.4$, 108 arcsec from,
and almost due north of, the target; the 
PS1 DR2 lists $g = 15.53, r = 14.98$ for this
star.  The light curves show a 
rise starting around phase zero, and a slower
decline, but no definite eclipse.  The spectroscopic
orbital period is not precise enough to specify phase
for the 2020 data; to prepare the figure, we assumed
the minimum around phase zero is stable in phase and
adjusted the period slightly to force its phase to align 
with the 2018 data.

\begin{figure}
\center
\includegraphics[width=5.0 in]{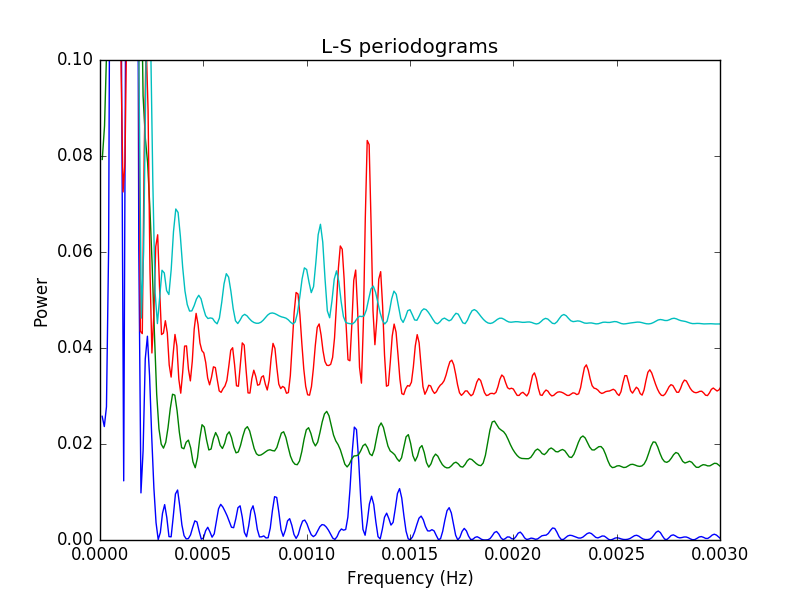}
\caption{Lomb-Scargle periodograms of the four 2018 light curves of SDSS1005+69, starting with the first night (26 Jan) at the bottom. QPOs are seen at $\sim$810 s on 26 Jan (blue) and 771 s on 1 Mar (red). 
}
\label{fig:sdss1005periodograms}
\end{figure}

\begin{figure}
\center
\includegraphics[width=5.0 in]{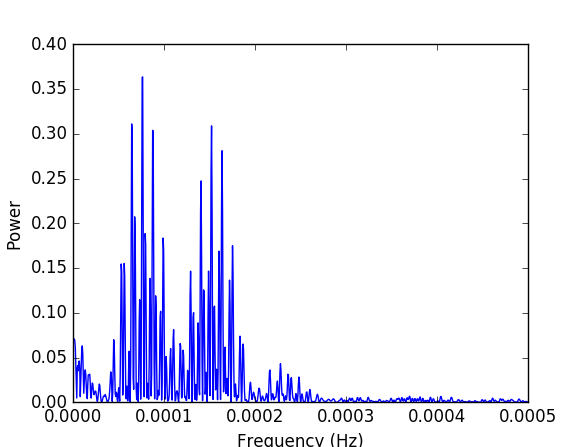}
\caption{Lomb-Scargle periodogram of the combined 2018 light curve of SDSS1005+69, showing the orbital frequency at 0.076 mHz (0.15 d) and its harmonic. The 1 cycle/d aliases are clearly seen. 
}
\label{fig:sdss1005LSorb}
\end{figure}

For the light curves taken on 26 February and 1 March 2018 (top panel of
Fig.~\ref{fig:sdss1005lightcurves}), we see evidence of possibly periodic
fluctuations on a $\sim$800 s timescales. We therefore produced periodograms of
the 2018 light curves using \textit{Gatspy}, a Python implementation of the
Lomb-Scargle method \citep{VanderPlas_2015}. The results are shown in
Fig.~\ref{fig:sdss1005periodograms}, which clearly show period peaks at 810 s
and 771 s, respectively, both of which have formal false-alarm
probabilities below 1 per cent. In addition, the periodogram of the four
combined nights clearly shows the presence of the orbital period and its
harmonic (Fig.~\ref{fig:sdss1005LSorb}). The fact that the two shorter period
peaks are not separated by the orbital frequency would seem to rule out an
intermediate polar interpretation, where the two frequencies could be due to
the beat and spin modulations, respectively. We therefore conclude that the
system probably exhibits quasi-periodic variability from time to time.   

The photometric variations seen in polars have variously been characterized as
flickering, fluctuations and sometimes quasi-periodic oscillations (QPOs). The
latter are variations that show some degree of coherence over a number of
cycles of the QPO period. The discovery of $\sim$Hz frequency QPOs in the
visible light of polars is now over 30 years old (e.g.
\citealt{1982ApJ...257L..71M}), and at the time resulted in a flurry of
theoretical studies. The commonly held understanding is that they are due to
plasma oscillations in the magnetically confined accretion columns. Until
recently, only five systems were known to exhibit such QPOs, with the sixth
(V379 Tel) being the only discovery in over two decades, despite attempts to
find more examples (Breytenbach, H. et al., in preparation).  Longer period
QPOs were also seen in AM Her \citep{1991A&A...251L..27B}, at 250$-$280s, while
more recently a $\sim$320 s QPO was detected in IGR J14536-5522 5.4
\citep{2010MNRAS.402.1161P}. The origin of these longer period QPOs is still
debated, with proposed sites suggested near the L1 point
\citep{1989MNRAS.241..365K}, the stream coupling region or within the
magnetically confined flow, close to the white dwarf surface
\citep{1991A&A...251L..27B}.

In polars the QPOs seem to occur with typically a few seconds period (accretion
column oscillations), or with periods of many minutes. The latter are larger
amplitude and are quite a common feature of polars, often seen by eye in the
lightcurves, as seems to be the case for SDSS J100516.61+694136.5 (see
Fig.~\ref{fig:sdss1005lightcurves}), where they can appear to show some sort of
coherence, but are not necessarily obvious in power spectra; they are often
referred to as ``QPO-like" \citep{2010MNRAS.402.1161P}.

\subsection{SDSS J133309.20+143706.9}
\label{subsec:sdss1333} 

\citet{schmidt08} published time-series spectroscopy and polarimetry 
of this object not long after it was discovered in SDSS.  The 
detection of circular polarization firmly established it as an
AM Her star.  The radial velocities of H$\alpha$ varied with 
$K \sim 250$ km s$^{-1}$ on a period of $2.2 \pm 0.1$ hr.
\citet{southworth15} obtained time-series photometry on
three nights, but were unable to improve on the period.

\begin{figure}
\includegraphics[width=6.5 in]{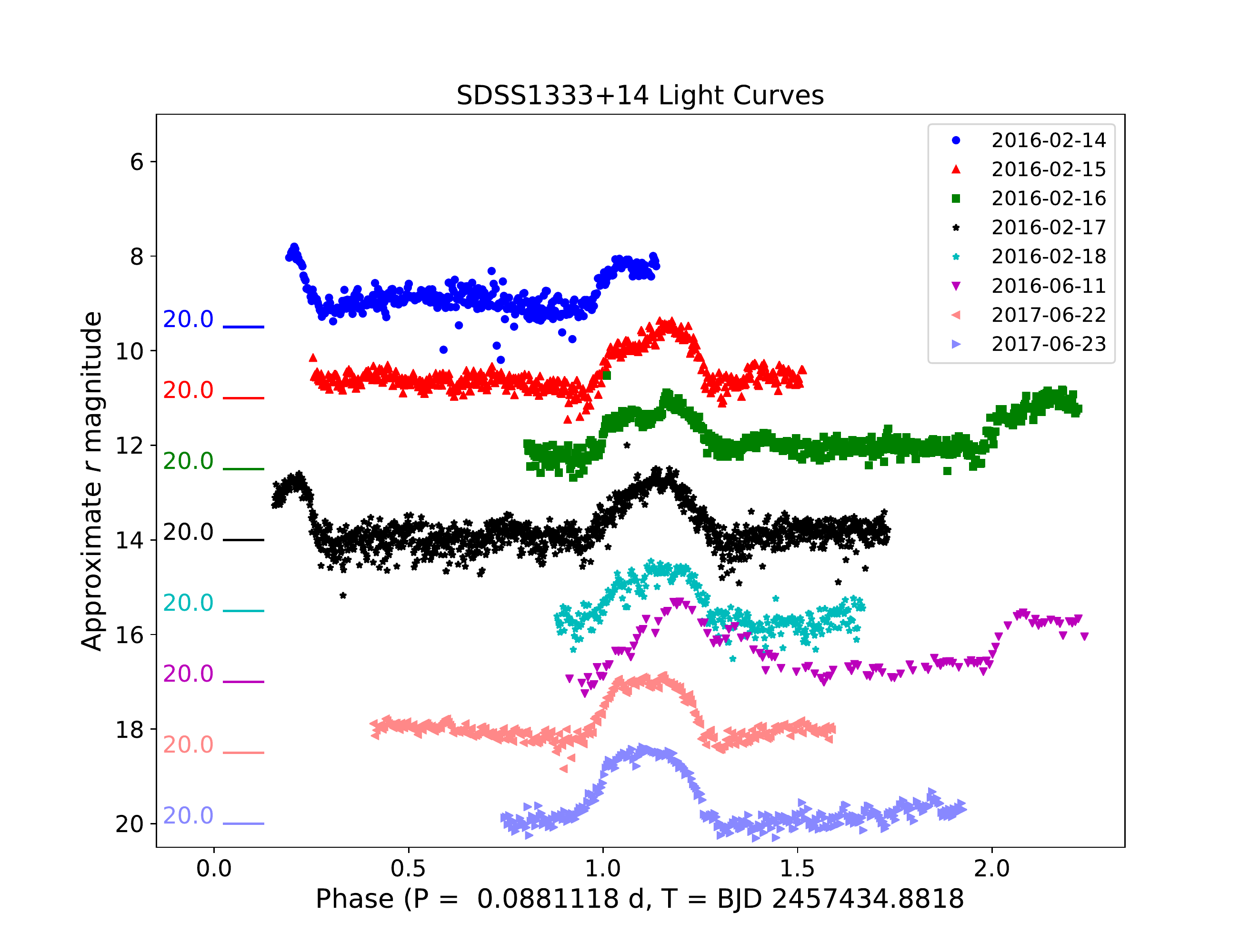}
\caption{Light curves of SDSS1333+14 from three observing runs.  
The differential magnitudes have been adjusted using the $r$ magnitudes
of the comparison stars.
The lowermost trace is plotted without a vertical offset,
and successive trace above it are offset upward by 1.5 mag, 
as indicated by the color-coded tick marks on the left.  
The ephemeris used to compute phases is
provisional, and is the epoch of the ingress into the 
bright phase.
}
\label{fig:sdss1333lightcurves}
\end{figure}

On 5 consecutive nights in 2016 February, we obtained time
series photometry with the 1.3m and Andor camera.  The light
curves (Fig.~\ref{fig:sdss1333lightcurves}) consistently
show a flat-topped brightening that 
recurs on a
period of 0.08814(4) d, or 126.92(6) min, consistent with 
the radial-velocity period found by \citet{schmidt08}.  
The daily cycle count is unambiguous.  

We also have time series from 2016 June 10 and 
from 2017 June 21 and 22.  The 2016 June time series shows a
brightening toward the end that is similar to those
seen in the other light curves, but does not
cover the decline. 
The 2017 June light curves show 
clearly-defined brightenings similar to the others.   
Only one choice of long-term cycle count fits all 
the brightening ingress times comfortably, and it implies
\begin{equation}
\label{eqn:sdss1333ephem}
\hbox{BJD of brightening} = 2457434.8817(4) + 0.0881118(3) E\ 
\ \hbox{(provisional).}
\end{equation}
We label this as provisional because of the lack of 
redundant timings on the longer baselines; the less
precise value from 2016 February is firmly established.
One reason for caution is that the three brightenings seen
in the \citet{southworth15} arrive early in this ephemeris 
by $\sim25$ min, in contrast to the MDM timings, which
all align to better than 2 min.

\subsection{SDSS J134441.83+204408.3}
\label{subsec:sdss1344}
\citet{szkody11} found 
this object (hereafter SDSS1344+20) in the SDSS data and noted
its apparently magnetic nature.  In a short series of spectra,
they found the radial velocities of H$\alpha$ and H$\beta$ varying
on a period of $\sim 115$ min, with semi-amplitudes $K \sim 400$
km s$^{-1}$.  \citet{szkody14} present further observations,
including photometry and spectroscopy showing changes of 
photometric state.  

\begin{figure}
\includegraphics[height=23 cm,trim = 1cm 4cm 0cm 5cm,clip=true]{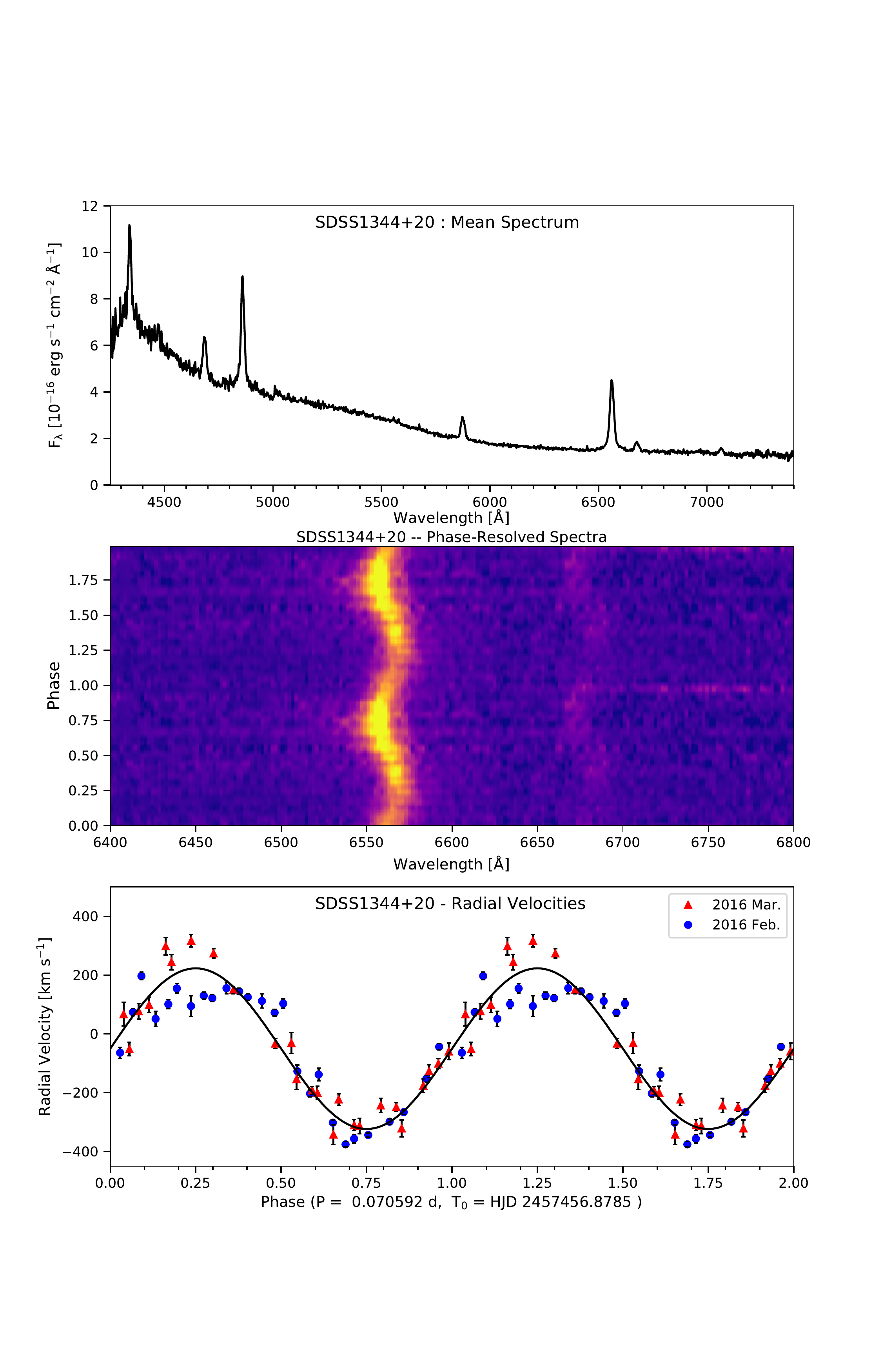}
\caption{Top:  Mean spectrum of SDSS1344+20 from 2016 February and March.  Middle: 
Phase resolved spectra of SDSS1344+20 in the region of 
H$\alpha$ and \ion{He}{1} $\lambda$6678.  Phase increases
from the bottom, and two cycles are shown for clarity.  Lower: 
Velocities of H$\alpha$, folded on the apparent orbital
period, with the best fit sinusoid superposed. 
}
\label{fig:sdss1344montage}
\end{figure}

We observed this star most intensively in 2016 February and 
March.  In the mean spectrum (Fig.~\ref{fig:sdss1344montage}),
\ion{He}{2} $\lambda 4686$ is less prominent that usual in 
AM Her stars, about half the strength of H$\beta$.  The 
continuum is strong and blue.  Hot continua usually show
a smooth upward sweep toward the blue; this continuum may have a very broad hump from $\sim$ 5100$-$5700\AA .  If real, this might be a 
cyclotron feature.  

As \citet{szkody11} found, the radial velocities of H$\alpha$ 
are strongly modulated, and with our more extensive data
set we determine $P_{\rm orb} = 101.652(6)$ min.  The cycle
count between nights and between the two observing runs is 
unambiguous; the relatively small uncertainty reflects the
21-day span of the time series.

\begin{figure}
\includegraphics[width=6.5 in]{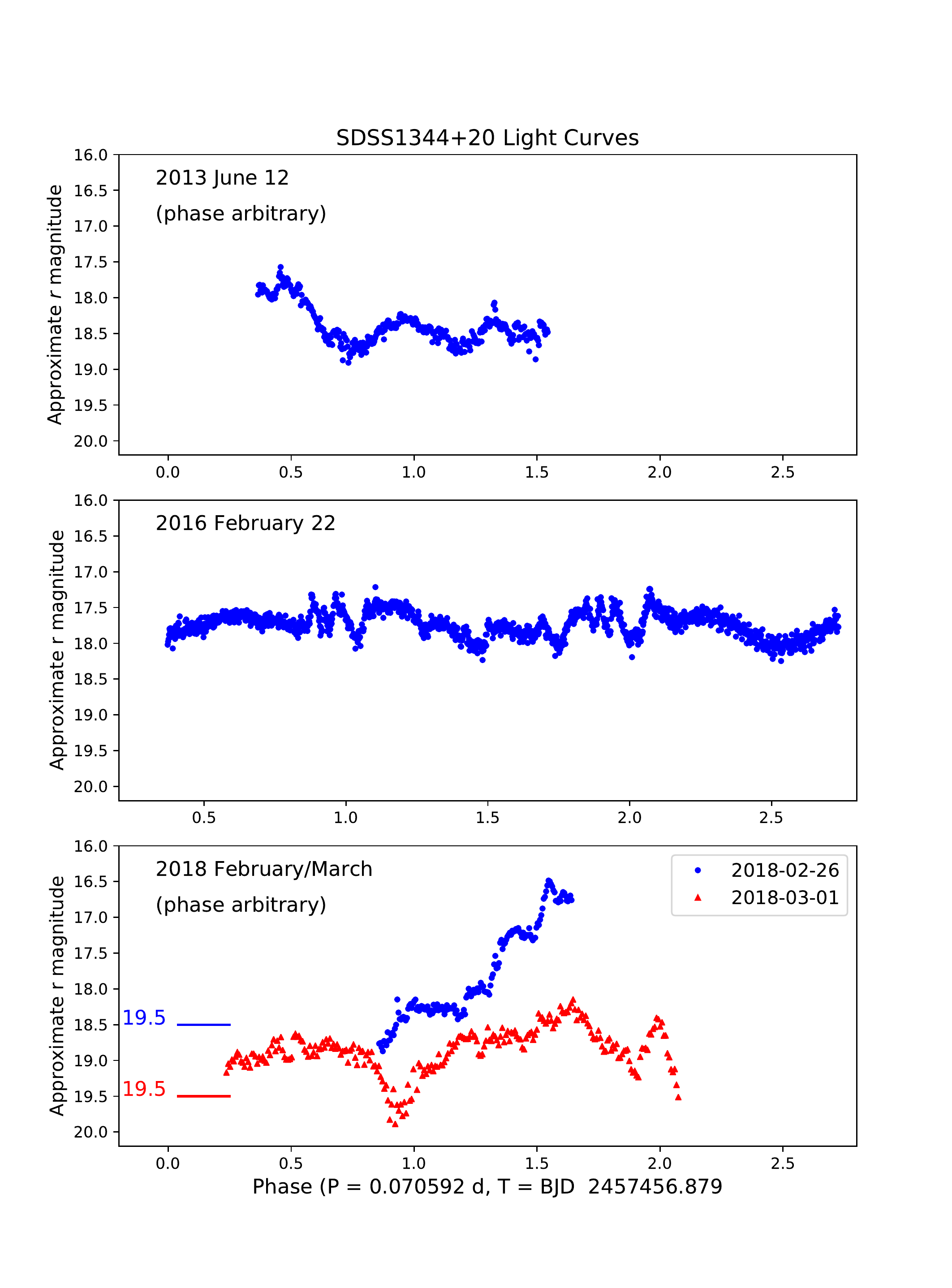}
\caption{Light curves of SDSS1344+20 from three observing runs.  
The differential magnitudes have been adjusted using the $r$ magnitudes
of the comparison stars.
The lower trace (red) in the lower panel is plotted without a vertical offset,
and the upper trace (blue) is offset upward by 1.0 mag.  The ephemeris
used to compute phase is given in the axis label, but it is
only valid for the middle panel, which was contemporaneous with 
the spectroscopy.
}
\label{fig:sdss1344lightcurves}
\end{figure}

Figure \ref{fig:sdss1344lightcurves} shows light curves
taken on three different observing runs; the spectroscopic
ephemeris used to compute the phases is only valid
for the 2016 February data, so the phases in the top
and bottom panels are arbitrary. 
Not all the runs used
the same comparison stars, but the magnitude scales
have been adjusted using the different stars' $r$ magnitudes
from PAN-STARRS.  No periodic 
behavior is evident, though the intervals of rapid
fluctuation seen in the middle panel are both centered
on a brief interval before phase zero.  
We speculate that the V-shaped $\sim$0.5 magnitude dips seen near phase zero in the 2016 data, when the object was brighter than for the other observations, could be a partial grazing eclipses of an accretion hot-spot.
The data from 2018-02-26 shows a brightening
by $\sim 2$ mag over less than one orbit; note that
\citet{szkody14} observed significant changes in the 
light curve from night to night.

\subsection{Gaia18aya}
\label{subsec:gaia18aya}

The Gaia light curve for this source shows it varying between
18 and 19 mag, except for a few days in 2018 April when it 
triggered an alert at a magnitude of 17.52, and a pair of
detections at 17.27 on 2018 May 25.    

\begin{figure}
\includegraphics[height=23 cm,trim = 1cm 4cm 0cm 5cm,clip=true]{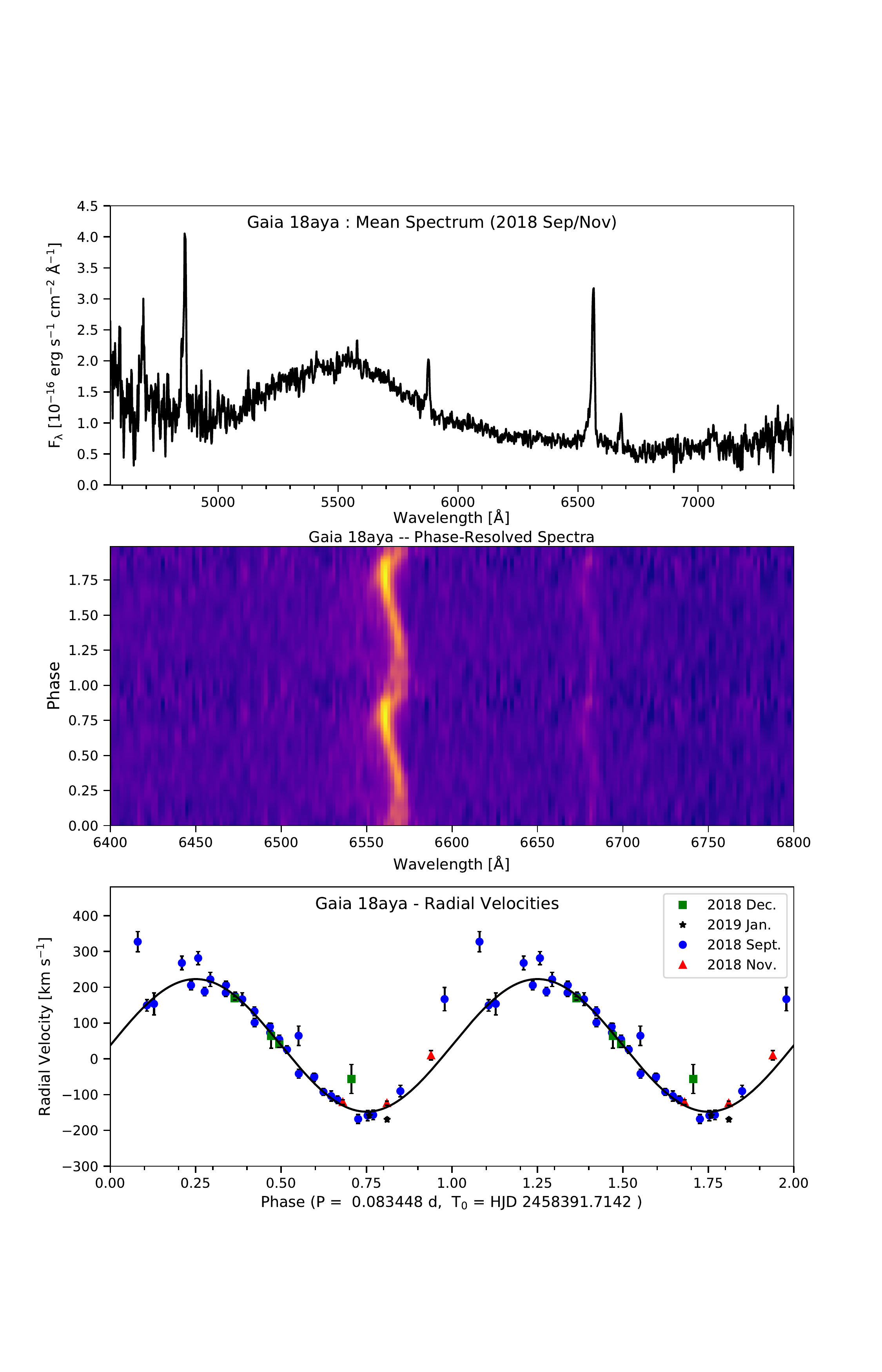}
\caption{
Top:  Mean spectrum of Gaia18aya from 2018 September and November, with the prominent cyclotron line at $\sim$5500 \AA.  Middle: 
Phase resolved spectra of Gaia18aya in the region of 
H$\alpha$ and \ion{He}{1} $\lambda$6678.  Phase increases
from the bottom, and two cycles are shown for clarity.  Lower: 
Velocities of H$\alpha$, folded on the apparent orbital
period, with the best fit sinusoid superposed. 
}

\label{fig:gaia18ayamontage}
\end{figure}

The mean spectrum (Fig.~\ref{fig:gaia18ayamontage}) shows
the usual emission lines, but the most striking feature is a 
cyclotron emission harmonic centered around $\sim 5500$ \AA .  
The cyclotron harmonic clinches the AM Her classification.  The 
H$\alpha$ radial velocities from 2018 September establish
an unambiguous orbital period near 120 min.  We obtained
more observations in 2018 November, December, and January
which constrain the period uniquely to 120.165(3) min. 

\begin{figure}
\includegraphics[width=7.0 in]{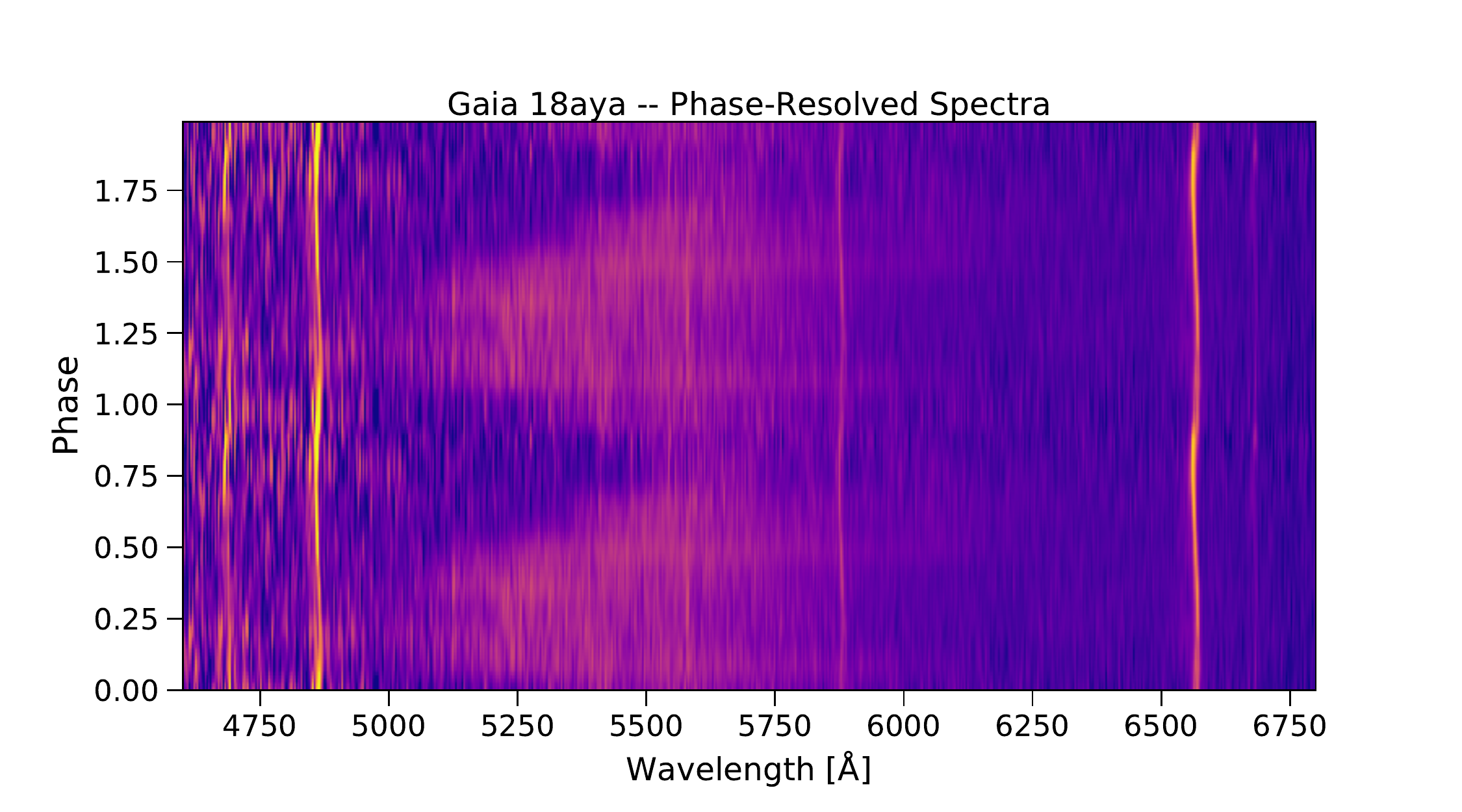}
\caption{Phase-resolved spectra of Gaia18aya, similar to the central 
panel of Fig.~\ref{fig:gaia18ayamontage}, but with a wider
wavelength range to show the phase dependence of the cyclotron
emission feature.}
\label{fig:gaia18ayawidetrail}
\end{figure}

The cyclotron emission hump varies in strength with the
orbital period.  This can be seen in Fig.~\ref{fig:gaia18ayawidetrail},
which is similar to the middle panel of Fig.~\ref{fig:gaia18ayamontage}
but with wider wavelength range.  In both these figures
the spectra were not rectified (normalized to a continuum)
before being averaged and stacked; rather, flux-calibrated
spectra were used, so variations in flux can be seen.

\begin{figure}
\includegraphics[width=6.5 in]{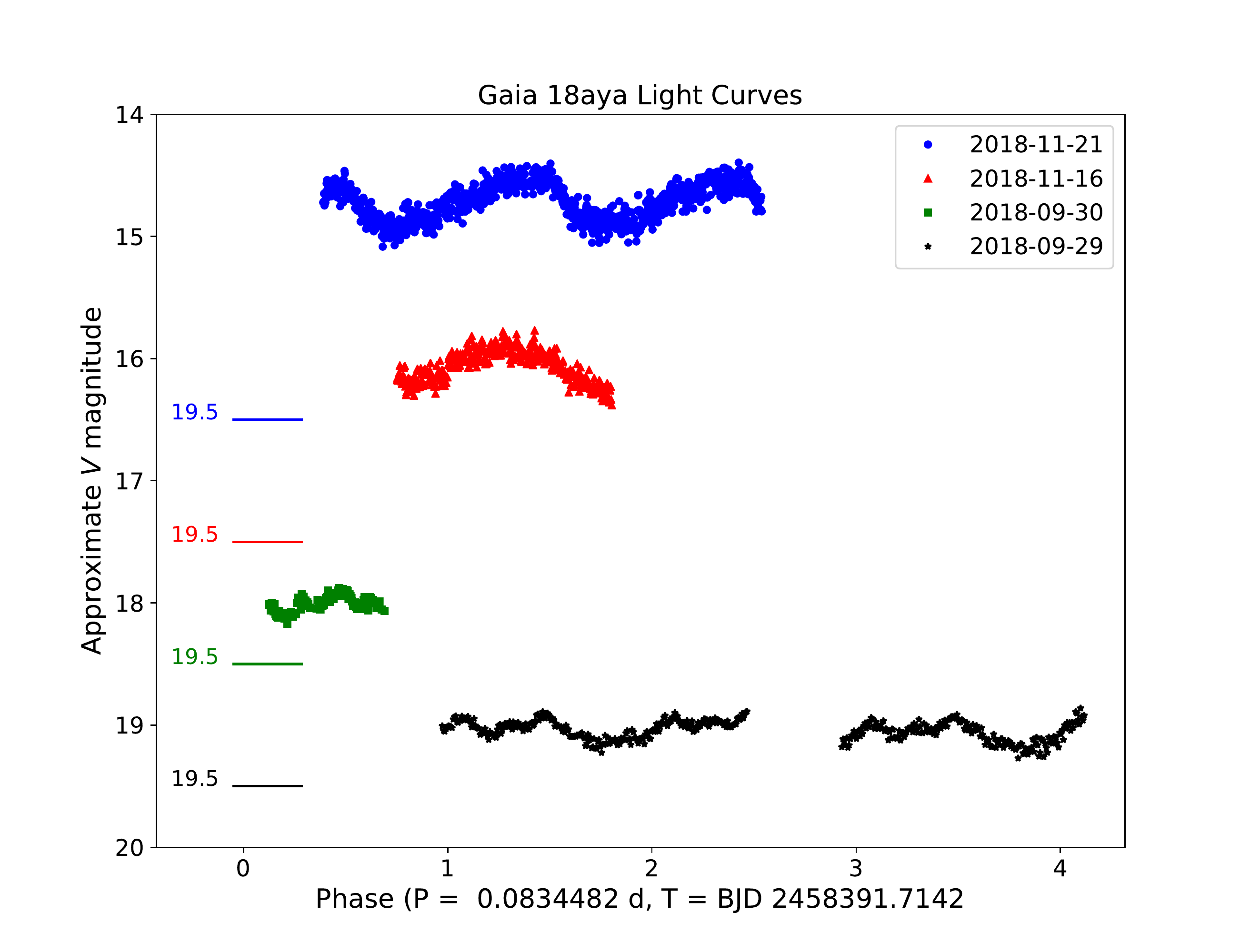}
\caption{Light curves of Gaia18aya from two observing runs.  
The differential magnitudes have been adjusted using the approximate
$V$ magnitudes of the comparison stars.
The lower trace is plotted without a vertical offset,
and the upper traces is offset upward by 1.5 mag, as indicate
by the color-coded tick marks.  
}
\label{fig:gaia18ayalightcurves}
\end{figure}

Fig.~\ref{fig:gaia18ayalightcurves} shows time series
photometry. During 2018 September, the variation is
irregular without obvious periodicity, but in 
2018 November the source was somewhat brighter
and varied smoothly with the orbital period. 

The wavelength of the $n$th cyclotron harmonic
is 
\begin{equation}
\label{eqn:cyclotronharmonic}
\lambda_n = {(10700\ \hbox{\normalfont{\AA}} ) \over n B_8},
\end{equation}
where $B_8$ is the magnetic field in units
of $10^8$ Gauss ($10^4$ Tesla).  The 
$\sim 5500$ \AA\ feature is the only harmonic
we clearly observe, which implies that the allowable magnetic fields for different assumed cyclotron harmonics in the range n = 2$-$7, varies from 28$-$97 MG. If the cyclotron feature at $\sim$5500 \AA\ is associated with the n = 6 (32 MG) or 7 (28 MG) harmonic, this implies that the shorter and longer wavelength harmonics at n $\pm$ 1 ( $\sim$4800 and $\sim$4700 \AA\ and $\sim$6400 and $\sim$6600 \AA\, respectively) should be detectable in our spectra. For n = 5 (B = 39 MG), we should also see the n = 4 harmonic at $\sim$6900 \AA. The fact that we see no other cyclotron features corresponding to these wavelengths is evidence that n < 5. If we take n = 4, then the neighbouring harmonics should occurs at $\sim$4400 \AA\ (n = 5) and $\sim$7300 \AA\ (n = 3), respectively. From Fig.~\ref{fig:gaia18ayamontage} (top panel) we see the flux increases from $\sim$7000 \AA\ to the red limit, at 7400 \AA\, consistent with a broad cyclotron line at $\sim$7300\AA. Similarly, the flux also increases for wavelengths $\leq$4900, to the blue limit of our spectra, at 4550 \AA\, also consistent with the expected cyclotron line at $\sim$4400 \AA. So this is all consistent with identifying the clearly observed hump at $\sim$5500 \AA\ with the n = 4 cyclotron harmonic from a B = 49 MG magnetic white dwarf. Lower harmonics, at n = 2 or 3, are also admittable, with higher field strengths, though the n $\pm$ 1 harmonics are now well outside the wavelength range of our spectra.

A good far-red or near-infrared spectrum could help determine the field strength by clearly identifying the lower harmnonics and allowing for subtraction of the underlying secondary star flux, which is likely an M-type star given the $\sim$2 h orbital period.

\section{Conclusions}
\label{sec:conclusions}

\begin{figure}
    \includegraphics{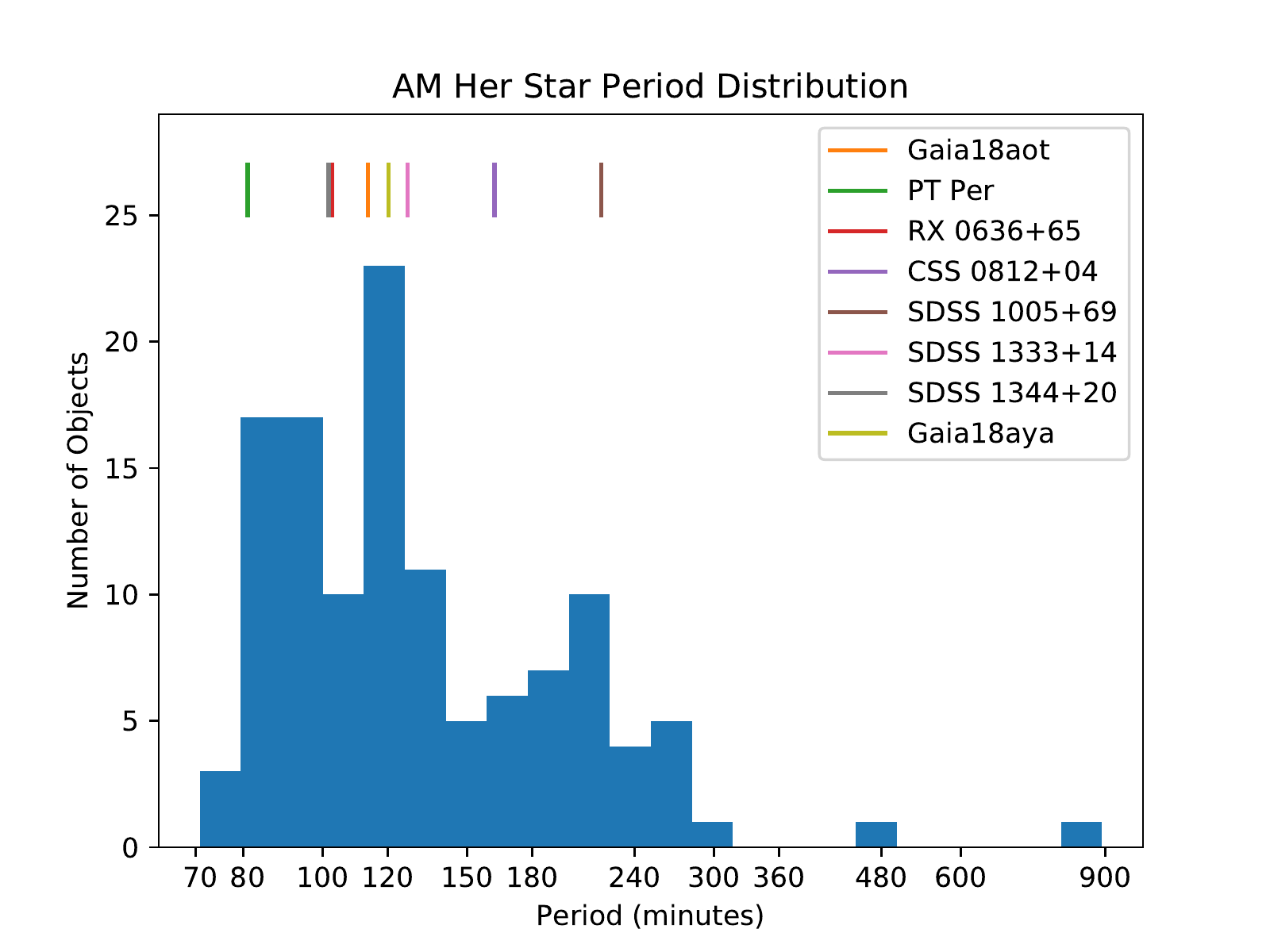}
    \caption{ Histogram of periods of AM Her stars listed in the final version (7.24) of the \citet{rkcat} catalog, along with tick marks indicating the periods of the present sample. The bins are uniform in $\log P$.  The tick marks for SDSS 1344+20 and RX 0636+65 nearly overlap.  The histogram point with the period near 15 hours (900 minutes) is V479 And, in which mass transfer may not occur through Roche-lobe overflow \citep{diego-gonzalez13}.}
    \label{fig:porbhist}
\end{figure}

Figure~\ref{fig:porbhist} shows a histogram of the orbital periods of AM Her
stars listed in the final release (version 7.24) of the \citet{rkcat} catalog of cataclysmic binaries.  The periods of the stars discussed here are also
indicated.  They all have periods typical of the population.

Table \ref{tab:summarytab} summarizes our findings.  We classify
three objects (Gaia18aot, RX J0636.3+6554, and Gaia18aya) 
as AM Her stars for the first time; the two Gaia sources are
also newly-recognized as CVs.
For six of the objects we 
determine $P_{\rm orb}$ for the first time, and for
two more (PT Per and SDSS1344+20) we improve significantly on
previous period determinations. We confirm that PT Per is a 
magnetic CV, as \citet{watson16} suggested.

Three of our objects have especially interesting light 
curves.  RX J0636.5+6554 eclipses deeply.  CSS0812+04 shows 
a sharp dip that is stable in phase and appears to 
be a partial eclipse.  Finally, SDSS1333+14 persistently
shows a distinctive bump consistent the the appearance
of an otherwise self-occulted accretion spot.

The spectrum of Gaia18aya has an apparent cyclotron
emission hump near 5500 \AA , which constrains the 
magnetic field to be greater than $\sim 49$ MG.  

It is worth noting that magnetic CVs appear to be
underrepresented in various listings. \citet{pala20}
constructed a volume-limited sample of 42 CVs 
within 150 pc, as judged by Gaia DR2 parallaxes, 
and found that over 30 per cent were magnetic, and 
that 11 out of the 42 in the total sample were polars, 
including the prototypical polar, AM Her.  High-cadence 
synoptic sky surveys have found very large numbers of new
CVS (see, e.g. \citealt{breedt14}), but they are
clearly biased toward dwarf novae, which show
distinct, large amplitude outbursts.  The objects in 
this paper no doubt represent a very sizeable population
of more subtly variable AM Her stars, as yet
unrecognized.

\begin{deluxetable}{llllrrl}
\label{tab:summarytab}
\tablewidth{0pt}
\tablecolumns{7}
\tablecaption{Summary of Findings}
\tablehead{
\colhead{Name} &
\colhead{Measurement} &
\colhead{$P$} &
\colhead{$T_0$} &
\colhead{$K$} &
\colhead{$\gamma$} &
\colhead{Remarks} \\ 
\colhead{} &
\colhead{} &
\colhead{[d]} &
\colhead{} &
\colhead{[km s$^{-1}$]} &
\colhead{[km s$^{-1}$]} &
\colhead{} \\
}
\startdata
Gaia18aot & Spec. &  0.078830(2) & 58443.7138(11) & 218(15)\tablenotemark{a} & $10(12)$ & New CV \\[1.2ex]
PT Per & Spec. & 0.05625(3) & 58504.7576(3) & 340(14) & $-9(9)$ & High state. \\[1.2ex]
RX J0636.3+6554 & Spec. & 0.07122(7) & 58174.6790(7) & 328(24)\tablenotemark{a} & $-12(14)$ & New AM Her \\
                & Eclipse & 0.071221298(8) & 58174.62129(5) & \nodata & \nodata & \\[1.2ex]
CSS080228:081210+040352  &  Spec. & 0.11247(17) & 57841.820(2) &  131(14) & $ 72(10)$ & New AM Her \\
                         & Dip & 0.11241902(2) &  57844.6379(2) & \nodata & \nodata & \\[1.2ex]
SDSS J100516.61+694136.5 & Spec. & 0.1518(3) & 58174.979(5) & 111(17)\tablenotemark{a} & $ 16(14)$ & New period \\[1.2ex]
SDSS J133309.20+143706.9 & Bump ingr. & 0.08812(4) & 57434.8816(6) & \nodata & \nodata & New per., firm \\
                         &            & 0.0881118(2) & 57434.8818(4) & \nodata & \nodata & Provisional \\[1.2 ex]
SDSS J134441.83+204408.3 & Spec. & 0.070592(4) & 57456.8785(7) & 273(17) & $-50(12)$ & Improved per. \\[1.2ex]
Gaia18aya & Spec. & 0.0834482(16) & 58391.7142(9) & 185(11) & $ 38(8)$ & New CV \\[1.2ex]
\enddata
\tablecomments{A summary of the measurements presented here.  Sinusoids,
where fitted, are of the form $v(t) = \gamma + K \sin{2 \pi (t - T_0) / P}$.
Epochs are barycentric Julian dates minus 2,400,000., in the UTC time
system; these can be converted to TDB with sufficient accuracy by 
adding 69 s.}
\tablenotetext{a}{Non-sinusoidal velocity curve; parameters are formal
best fits only.}
\end{deluxetable}

\acknowledgments
The U.S. National Science Foundation supported a portion of this work through grant AST-1008217.
We thank Karolina B\k{a}kowska for obtaining some MDM
time series data that are used here.
The CSS survey is funded by the National Aeronautics and Space
Administration under Grant No.~NNG05GF22G issued through the Science
Mission Directorate Near-Earth Objects Observations Program.  The CRTS
survey is supported by the U.S.~National Science Foundation under
grants AST-0909182 and AST-1313422.
DB acknowledges research support from the South African National Research Foundation. 
We acknowledge ESA Gaia, DPAC and the Photometric Science 
Alerts Team (http://gsaweb.ast.cam.ac.uk/alerts).
We thank the MDM Observatory staff for their
tireless efforts to keep the telescopes on sky.
Our heartfelt thanks go to the
Tohono O'odham Nation for letting us use their mountain
to explore the sky that surrounds us all.

\bibliographystyle{yahapj}
\bibliography{ref}

\end{document}